\documentclass[prd,twocolumn,superscriptaddress,amsfonts,amssymb,amsmath,showpacs]{revtex4-2}
\usepackage{bm}
\usepackage{latexsym}
\usepackage[latin1]{inputenc}
\usepackage{graphicx}
\usepackage{amsmath}
\usepackage{amssymb}
\usepackage{amsthm}
\usepackage{relsize}
\usepackage{palatino}
\usepackage{mathpazo}
\usepackage{textcomp}
\usepackage{float}
\usepackage{booktabs}
\usepackage{dcolumn}
\usepackage{booktabs}
\usepackage{multirow}
\usepackage{tikz}
\usepackage{hyperref}
\hypersetup{
    colorlinks=true,
    linkcolor=purple,
    filecolor=magenta,      
    citecolor=blue
}
\usepackage{amsmath}
\usepackage{xcolor}
\usepackage{orcidlink}
\usepackage{graphicx,epsfig}
\usepackage{graphicx,subfigure}
\usepackage{commath}

\makeatletter
\long\def\dddddot#1{%
  {\mathop {#1}\limits ^{\vbox to-1.4\ex@ {\kern -\tw@ \ex@ \hbox {\normalfont .....}\vss }}}%
}
\long\def\multidots#1#2{%
  \count@=0
  {{\mathop {#2}\limits ^{\vbox to-1.4\ex@ {\kern -\tw@ \ex@ \hbox {\normalfont %
  \loop%
  \ifnum#1>\count@%
  .%
  \advance\count@ by1%
  \repeat%
  }\vss }}}}%
}
\makeatother

\begin{document}

\title{\bf Non-Metricity Corrections Approach to Alleviate $H _0$ Tension: The Logarithmic and Nonlinear $f(Q)$ Models}

\author{Rahul Bhagat\orcidlink{0009-0001-9783-9317}}
\email{rahulbhagat0994@gmail.com}
\affiliation{Department of Mathematics, Birla Institute of Technology and Science, Pilani, Hyderabad Campus, Jawahar Nagar, Kapra Mandal, Medchal District, Telangana 500078, India.}
\author{Kesava Chodavarapu\orcidlink{0000-0000-0000-0000}}
\email{ckesavaswarup@gmail.com}
\affiliation{Department of Mathematics, Birla Institute of Technology and Science, Pilani, Hyderabad Campus, Jawahar Nagar, Kapra Mandal, Medchal District, Telangana 500078, India.}
\author{B. Mishra\orcidlink{0000-0001-5527-3565}}
\email{bivu@hyderabad.bits-pilani.ac.in}
\affiliation{Department of Mathematics, Birla Institute of Technology and Science, Pilani, Hyderabad Campus, Jawahar Nagar, Kapra Mandal, Medchal District, Telangana 500078, India.}

\begin{abstract}
The persistent discrepancy between early-time and late-Universe measurements of the Hubble constant commonly known as the $H_0$ tension remains one of the most pressing open questions in modern cosmology. In this work, we explore whether modifications to the gravitational sector, specifically 
within the framework of symmetric teleparallel gravity, can offer a viable 
pathway toward alleviating this tension. We consider two functional forms 
of $f(Q)$ gravity: a logarithmic model and a nonlinear saturation model, 
both of which introduce geometric corrections to the standard expansion 
history without invoking a cosmological constant. Constraining these models 
through a Bayesian MCMC analysis against a comprehensive suite of 
observational data, including CC, Type Ia supernovae 
compilations (Pantheon, Pantheon$+$SH0ES, and DES SN5YR), and BAO 
measurements from SDSS and DESI, we find that both models remain 
statistically competitive with $\Lambda$CDM. The logarithmic model yields $H_0$ values that reduce the statistical discrepancy with the SH0ES determination for the considered late-time dataset combinations, with the residual tension quantified using a standard $\sigma$-based tension metric. Although the AIC, BIC, and DIC information criteria marginally prefer $\Lambda$CDM over the extended models, the results indicate that the logarithmic model can partially alleviate the $H_0$ tension for the late-time datasets considered, with the degree of alleviation depending on the specific combination of observations, providing a phenomenological assessment of late-time $H_0$ tension within the modified gravity framework.
\end{abstract}

\maketitle

\section{Introduction}

The Hubble constant $H_0$, which quantifies the present expansion rate of the Universe, has now been measured with remarkable precision, transforming what was once a minor inconsistency into a significant challenge for the standard cosmological model. This discrepancy, commonly referred to as the \textit{Hubble tension}~\cite{DiValentino_2021_131}, corresponds to a persistent disagreement-now exceeding the $5\sigma$ level-between values of $H_0$ inferred from different observational probes~\cite{n86r-sjgm,Jia_2025,CosmoVerseNetwork:2025alb}. The discovery of the late-time accelerated expansion of the Universe through observations of Type Ia supernovae (SNIa)~\cite{Riess_1998_116,Perlmutter_1998_517} established these objects as powerful distance indicators, forming the basis of the cosmic distance ladder. In particular, late-Universe measurements calibrated via Cepheid variables and SNIa consistently yield a higher value, $H_0 = 73.04 \pm 1.04\ \mathrm{km\ s^{-1}\ Mpc^{-1}}$~\cite{Riess_2021_934}, further supported by recent analyses incorporating \textit{James Webb Space Telescope} (JWST) observations \cite{Riess:2025chq} and extended distance ladder studies~\cite{H0DN:2025lyy}. In contrast, baryon acoustic oscillation (BAO) measurements provide an independent probe of the expansion history by utilizing the sound horizon at the baryon drag epoch, $r_d \approx 147\,\mathrm{Mpc}$, as a standard ruler~\cite{Conley_2010}. This scale, imprinted in the large-scale distribution of matter and observed in the galaxy two-point correlation function, enables precise constraints on both the comoving angular diameter distance $D_M(z)$ and the Hubble parameter $H(z)$ over a wide redshift range. Large spectroscopic surveys such as WiggleZ, SDSS-III BOSS~\cite{Eisenstein_2005_633,Alam_2017_470,Raichoor_2020_500}, and more recently the Dark Energy Spectroscopic Instrument (DESI) have significantly improved the precision of BAO measurements, with the latest DESI Data Release 2 (DR2), covering $0.1 < z < 4.2$, reporting $H_0 = 68.51 \pm 0.58\ \mathrm{km\ s^{-1}\ Mpc^{-1}}$~\cite{DESI:2025zgx,Hou_2020_500}, in closer agreement with early-Universe estimates. The contrast between the higher values inferred from supernova-based distance ladder measurements and the lower values obtained from BAO lies at the core of the Hubble tension. 

More recently, this growing inconsistency has placed increasing pressure on the standard $\Lambda$CDM framework~\cite{DIVALENTINO2021102605}, motivating a systematic re-examination of its underlying assumptions and possible extensions. A broad class of theoretical approaches has been proposed to address this tension, particularly those that modify the late-time expansion history $H(z)$ at redshifts $z \lesssim 2$~\cite{Cheng_2025,DIVALENTINO2020100666,qqc6-76z4}. Among these, dynamical dark energy models replace the cosmological constant with a time-dependent component characterized by an evolving equation of state parameter $w = p/\rho$, thereby introducing additional degrees of freedom that must account for nearly 70 percent of the present energy budget of the Universe~\cite{kjpb-r698,Lobo_2009_173,Nojiri_2006_74_086005,Nojiri_2005_631,S0219887824501330}. While such models raise important questions regarding the physical origin and non-detection of these fields, recent observational developments have renewed interest in these scenarios. In particular, the latest BAO measurements from the DESI collaboration \cite{DESI:2025zgx,qqc6-76z4,Jia_2025,PALIATHANASIS2025101993,dj3k-84v4}, extending earlier results from SDSS surveys, have significantly enhanced the precision of cosmological constraints. Complementary information is provided by Type Ia supernova datasets, including the Pantheon (PN) and Pantheon+SH0ES compilations as well as the Dark Energy Survey five-year sample (DES SN5YR), which offer improved control over systematic uncertainties~\cite{Scolnic_2018_859,Brout_2022_938,Scolnic_2022,Lopez-Hernandez:2025lbj,Abbott_2018_480,DEScollaboration2025}. When combined with local measurements from the SH0ES program, these datasets consistently favor a higher value of $H_0$~\cite{Riess_2022}. Within this observational context, recent Bayesian analyses-especially those incorporating DESI data-have indicated a statistical preference for dynamical dark energy models over the standard $\Lambda$CDM scenario~\cite{CosmoVerseNetwork:2025alb,10.1093/mnrasl/slaf042,Sultana_2025,S0218271824500354}, even after accounting for the presence of additional free parameters. This emerging preference suggests that deviations from a simple cosmological constant ($w = -1$) may be favored, potentially providing new insights into the nature of cosmic acceleration and its connection to the Hubble tension~\cite{Stern_2010_2010_008,Esculpi_2010_67,Gruber_2014_89,Aviles_2012_86,DEVI2026170597}.

An alternative and conceptually distinct approach involves modifying the underlying theory of gravity rather than introducing new energy components. In this context, cosmic acceleration arises from deviations from General Relativity (GR) on large scales, leading to modifications in the expansion history and the inferred cosmological parameters \cite{Sotiriou_2008_82,Farrugia_2016_93,S0219887822501791,JAVED2024101656,kala2025,Ashraf2026,Ashraf12026,ASHRAF2026102274,Tillayev2026}. Such theories can, in principle, influence both early and late-time physics either by altering the sound horizon scale or by modifying the late-time dynamics thereby offering additional flexibility in addressing the Hubble tension. Among the various modified gravity frameworks, symmetric teleparallel gravity, or $f(Q)$ gravity, has recently gained significant attention~\cite{Jimenez_2018_98}. In this formulation, gravity is described in terms of the non-metricity scalar $Q$, and the gravitational action is generalized to an arbitrary function $f(Q)$, introducing new dynamical degrees of freedom capable of driving cosmic acceleration without invoking a cosmological constant~\cite{Jimenez_2020_101,Bhgat_2026_Adf,Anagnostopoulos_2021_822,Barros_2020_30,Bajardi_2020_135,Khyllep_2021_103,BHAGAT_JHEAP3}. The $f(Q)$ framework is particularly appealing in the context of the Hubble tension, as it modifies the effective Friedmann equations and can produce deviations in the late-time expansion rate while preserving the well-established predictions of early-Universe cosmology\cite{Paliathanasis_2023_41,SULTANA2026100620,Capozziello_2022_832,Bhagat_2026_adf2}. Consequently, these models can effectively mimic dynamical dark energy behavior, including a redshift-dependent equation of state, and offer a viable pathway toward reconciling the discrepancy in $H_0$~\cite{Lazkoz_2019_100,SULTANA2025100422,Bohmer_2023_9,Flathmann_2021_103,Maurya_2022_2022_003,BHAGAT_2025_new}. Recent studies show that EDE motivated by the Hubble tension can increase the inferred \(\alpha_s\), strengthening tension with USR-based PBH formation \cite{sabogal2026}. In another model-independent test, the detection of a Cosmic Graviton Background (\(\Delta N_{\rm eff,g}\simeq0.054\)) could rule out realistic inflationary scenarios, offering a potential probe of alternatives such as bouncing and emergent cosmologies \cite{Vagnozzi2022,HEISENBERG20241}. Confronting such models with recent observational datasets including BAO measurements from DESI, supernova compilations such as Pantheon+ and DES SN5YR, and local determinations of the Hubble constant provides a crucial test of whether modifications to gravity can alleviate the tension and yield an improved description of the observed Universe relative to the standard $\Lambda$CDM model~\cite{Freedman_2021_919,Vagnozzi_2022,DiValentino_2021_131,Abdalla_2022_34,Riess_2021_908,DIVALENTINO2020100666,Jia_2025}.

In this work, we investigate the implications of symmetric teleparallel gravity within the context of the Hubble tension by $f(Q)$ model with a comprehensive set of late-time observational data. In particular, we consider multiple Type Ia supernova compilations, including the Pantheon, Pantheon+ combined with SH0ES, and the DES SN5YR, which together provide robust constraints on the expansion history over a wide redshift range. These datasets are further complemented by BAO measurements from SDSS and the latest results from the DESI, as well as cosmic chronometer (CC) data that offer direct estimates of the Hubble parameter. The primary objective of this analysis is to examine how modifications arising from symmetric teleparallel gravity, specifically within the framework of $f(Q)$ theories, influence the late-time expansion dynamics and whether they can alleviate the existing tension in $H_0$ measurements. To this end, we consider two representative functional forms of $f(Q)$: a logarithmic model and a nonlinear saturation model, both of which introduce deviations from the standard $\Lambda$CDM behavior through geometric effects.

The structure of the paper is organized as follows. In Section~\ref{Sec:2}, we present the theoretical framework of $f(Q)$ gravity and derive the corresponding cosmological field equations for a homogeneous and isotropic Universe. Section~\ref{Sec:3} describes the observational datasets employed in this study and outlines the statistical methodology used for parameter estimation. In Section~\ref{Sec:4}, we introduce the specific functional forms of the proposed models and derive the associated expressions for the Hubble expansion rate. This section also presents the observational constraints obtained within a Bayesian framework using the combined datasets, together with a comparative analysis of Model~I and Model~II. Finally, in Section~\ref{Sec:5}, we summarize the main results and discuss their implications for the Hubble tension, as well as the viability of modified gravity as an alternative to the standard cosmological model.\\

\section{Field Equations of $f(Q)$ gravity}\label{Sec:2}

The action for $f(Q)$ gravity \cite{Jimenez_2018_98} is given by
\begin{equation}\label{Eq:1}
    S = \frac{1}{2} \int d^4x \sqrt{-g} f(Q) + S_m,
\end{equation}
where we adopt natural units such that $8\pi G/c^4 = 1$, and $S_m$ denotes the matter action. By varying the action with respect to the metric tensor, one obtains the field equations as
\begin{multline}\label{Eq:2}
      \frac{2}{\sqrt{-g}}\nabla_\alpha \left( \sqrt{-g} f_Q P^\alpha_{~~\mu\nu} \right) - \frac{1}{2} f g_{\mu\nu} \\+ f_Q \left( P_{\mu\alpha\beta} Q_{\nu}^{~~\alpha\beta} - 2  Q^{\alpha\beta}_{~~~\mu}P_{\alpha\beta\nu} \right) = T_{\mu\nu},
\end{multline}
where $ f_Q = \partial f/\partial Q $. In addition, variation with respect to the affine connection leads to
\begin{equation}\label{Eq:3}
    \nabla_\mu \nabla_\nu \left( \sqrt{-g} f_Q P^{\mu\nu}_{~~~\alpha}\right) = 0,
\end{equation}
which governs the dynamics associated with the connection.

The tensor $ P^\alpha_{\ \mu\nu} $, conjugate to the nonmetricity, is defined as
\begin{equation}\label{Eq:4}
    P^{\alpha}_{~~\mu\nu} = -\frac{1}{4} Q^{\alpha}_{\mu\nu} + \frac{1}{2} Q_{(\mu\ \nu)}^{\ \ \ \alpha} + \frac{1}{4} (Q^\alpha - \tilde{Q}^\alpha) g_{\mu\nu} - \frac{1}{4} \delta^{\alpha}_{~(\mu} Q_{\nu)},
\end{equation}
where parentheses denote symmetrization, $ A_{(\mu\nu)} = \frac{1}{2}(A_{\mu\nu} + A_{\nu\mu}) $. The two independent traces of the nonmetricity tensor are defined as
\begin{equation}\label{Eq:5}
    Q_\alpha = g^{\sigma\lambda} Q_{\alpha\sigma\lambda}, \quad \tilde{Q}_\alpha = g^{\sigma\lambda} Q_{\sigma\alpha\lambda}.
\end{equation}
The nonmetricity scalar $Q$ can then be expressed as
\begin{equation}\label{Eq:6}
Q = -Q_{\alpha\mu\nu} P^{\alpha\mu\nu}.
\end{equation}

To study cosmological implications, we consider a spatially flat, homogeneous, and isotropic Friedmann-Lema\^itre-Robertson-Walker (FLRW) spacetime described by
\begin{equation}\label{Eq:7}
    ds^2 = -dt^2 + a^2(t) \left(dx^2 + dy^2 + dz^2\right),
\end{equation}
where $a(t)$ is the scale factor. In the coincident gauge, the connection vanishes and the metric becomes the only dynamical variable. Choosing a different gauge would introduce nontrivial contributions from the connection \cite{Jimenez_2018_98,HEISENBERG20241}. For the metric given in Eq.~\eqref{Eq:7}, the nonmetricity scalar reduces to
\begin{equation}\label{Eq:8}
Q = 6H^2,
\end{equation}
where $H = \dot{a}/a$ is the Hubble parameter.

The matter content of the Universe is assumed to be a perfect fluid with energy-momentum tensor
\begin{equation}\label{Eq:9}
T_{\mu\nu} = (\rho + p)u_\mu u_\nu + p g_{\mu\nu},
\end{equation}
where $\rho$ and $p$ denote the energy density and pressure, respectively, and $u_\mu$ is the four-velocity of the fluid. Substituting the FLRW metric into the field equations leads to the modified Friedmann equations
\begin{eqnarray}\label{Eq:10}
6H^2 f_Q - \frac{1}{2} f &=& \rho, \\
(12H^2 f_{QQ} + f_Q)\dot{H} &=& -\frac{1}{2}(\rho + p). \label{Eq:11}
\end{eqnarray}
These equations determine the cosmological dynamics in $f(Q)$ gravity.

By decomposing the function as $f(Q) = Q + \Phi(Q)$, the above equations can be rewritten as
\begin{equation}\label{Eq:12}
     3H^2 = \rho + \frac{1}{2}\Phi - Q \Phi_Q,
\end{equation}
\begin{equation}\label{Eq:13}
     -3H^2 - 2\dot{H} = p + Q \Phi_Q - \frac{1}{2}\Phi + 2\dot{H}(2Q \Phi_{QQ} + \Phi_Q).
\end{equation}
To construct a viable cosmological model, it is necessary to specify a functional form for $\Phi(Q)$.

\section{Observational Data Sets and Statistical Methodology}\label{Sec:3}

To constrain the model parameters, we perform a Bayesian analysis using a Markov Chain Monte Carlo (MCMC) approach implemented via the publicly available \texttt{emcee} package~\cite{Foreman-Mackey_2013_125}, which utilizes an affine-invariant ensemble sampler. This algorithm is well suited for efficiently exploring correlated, high-dimensional parameter spaces. The posterior probability distribution is constructed as $\mathcal{P}(\Theta|D) \propto \mathcal{L}(D|\Theta)\,\Pi(\Theta)$, where $\mathcal{L}$ denotes the likelihood function and $\Pi(\Theta)$ represents the prior on the parameter vector $\Theta$. Assuming Gaussian-distributed observational uncertainties, the total likelihood is expressed in terms of the combined chi-square as
\begin{equation}
\chi^2_{\text{tot}}(\Theta) = \chi^2_{CC} + \chi^2_{\text{SN}} + \chi^2_{\text{BAO}}.
\end{equation}
We evolve multiple chains (walkers) in parallel, ensuring proper sampling of the posterior distribution. Convergence is assessed through standard diagnostics such as the stabilization of chain means and the integrated autocorrelation time, and only samples from the converged, post-burn-in phase are used to derive parameter constraints.

\vspace{0.3cm}

\textbf{Cosmic Chronometers (CC):}  
We utilize a compilation of 31 measurements of the Hubble parameter obtained via the CC technique \cite{Moresco_2015_450,Moresco_2022_25}. This method provides direct estimates of $H(z)$ by measuring the differential age evolution of passively evolving galaxies, exploiting the relation $H(z) = -\frac{1}{1+z}\frac{dz}{dt}$. Since it relies on relative age differences $\Delta z/\Delta t$, it is less sensitive to systematic uncertainties compared to methods based on absolute age determinations \cite{Stern_2010_2010_008,Borghi_2022_928,Bhagat_ASPdyna2024}. The estimation of galaxy ages is based on well-calibrated stellar population synthesis models \cite{Jimenez_2002,Moresco_2012_2012_006,Sultana:2026ych,BHAGAT2025101913}, and the method remains largely independent of any assumed cosmological model or distance ladder calibration. The corresponding chi-square function is given by
\begin{equation}
\chi^2_{CC}(\Theta) = \sum_{i=1}^{31} \frac{\left[H(z_i,\Theta) - H_{\text{obs}}(z_i)\right]^2}{\sigma_{CC}^2(z_i)},
\end{equation}
where $H(z_i,\Theta)$ and $H_{\text{obs}}(z_i)$ denote the theoretical and observed values, respectively, with $\sigma_{CC}(z_i)$ representing the measurement uncertainty.

\vspace{0.3cm}

\textbf{Type Ia Supernovae:}  
Another key dataset employed in our MCMC analysis consists of observations from Type Ia supernovae (SNe Ia)~\cite{Riess_1998_116,Perlmutter_1998_517}. These events, typically originating from thermonuclear explosions in binary systems, are of particular importance in cosmology due to their nearly uniform intrinsic luminosity. This property allows them to be used as standardizable candles for measuring distances to high-redshift galaxies and thereby constraining the expansion history of the Universe.

The primary observable in supernova cosmology is the distance modulus, defined as the difference between the apparent magnitude \(m\) and the absolute magnitude \(M\)~\cite{Phillips_1993_413}, given by
\begin{equation}
\mu(z_i,\Theta) = m - M = 5 \log_{10}\left[D_L(z_i,\Theta)\right] + 25,
\end{equation}
where the luminosity distance is expressed as
\begin{equation}
D_L(z_i,\Theta) = c(1+z_i)\int_0^{z_i} \frac{dz'}{H(z',\Theta)}.
\end{equation}
Here, \(c\) denotes the speed of light and \(H(z,\Theta)\) is the model-dependent Hubble parameter.

In this work, we utilize two major SNe Ia compilations: the Pantheon dataset and its extended version, Pantheon+ combined with SH0ES~\cite{Riess_2021_934,Riess_2022,Scolnic_2022,Brout_2022_938}. The Pantheon sample consists of 1048 spectroscopically confirmed supernovae spanning the redshift range \(0.01 < z \lesssim 2.3\), providing a robust baseline for cosmological analysis~\cite{Najera_2021_34,Scolnic_2018_859}. The Pantheon+SH0ES compilation significantly extends this dataset to 1701 light-curve measurements from 1550 confirmed SNe Ia \cite{Riess_2018_853,Amanullah_2010_716,Suzuki_2012_746,Conley_2010}, incorporating Cepheid-calibrated host galaxy distances from the SH0ES program. This calibration plays a crucial role in breaking the degeneracy between the absolute magnitude \(M\) and the Hubble constant \(H_0\), thereby enabling tighter constraints on cosmological parameters. 

For both the Pantheon and Pantheon+SH0ES datasets, the absolute magnitude \(M\) is treated as a nuisance parameter and is marginalized over during the MCMC analysis. The theoretical predictions for the distance modulus are compared with observations, and parameter estimation is performed by minimizing the corresponding chi-square function,
\begin{equation}
\chi^2_{\text{SN}} = \Delta \mu^T C^{-1} \Delta \mu,
\end{equation}
where \(\Delta \mu = \mu(z_i,\Theta) - \mu_{\text{obs}}(z_i)\), and \(C\) denotes the covariance matrix that includes both statistical and systematic uncertainties.

{ In addition to these datasets, we also consider the Dark Energy Survey five-year supernova sample (DES-SN5YR), which comprises 1829 SNe Ia and provides an independent and homogeneous dataset with improved control over observational systematics~\cite{DEScollaboration2025,Lopez-Hernandez:2025lbj,Abbott_2018_480}. We use the publicly released DES-SN5YR distance-modulus data together with the corresponding full statistical and systematic covariance matrix. The published distance moduli are constructed using a reference normalization of $H_0$ stated Ref~\cite{10.1093/mnras/staf122}. This reference value is part of the normalization adopted for the released distance-modulus data and is not imposed as a fixed value or external prior on $H_0$ in our analysis.

Since the unanchored DES-SN5YR Hubble diagram does not independently determine the absolute supernova luminosity, the absolute magnitude and $H_0$ remain degenerate~\cite{DEScollaboration2025}. Therefore, rather than fixing the absolute magnitude $M$, we account for the corresponding additive calibration uncertainty by introducing a nuisance offset $\mathcal{M}$ and analytically marginalizing over it. In particular, defining $\Delta\boldsymbol{\mu}=\boldsymbol{\mu}_{\rm obs}-\boldsymbol{\mu}_{\rm th}$, we use
\begin{equation}
\chi^2_{\rm DES}(\mathcal{M}) =
(\Delta\boldsymbol{\mu}-\mathcal{M}\mathbf{1})^{T}
\mathbf{C}_{\rm DES}^{-1}
(\Delta\boldsymbol{\mu}-\mathcal{M}\mathbf{1}),
\end{equation}
where $\mathbf{C}_{\rm DES}$ is the full DES-SN5YR covariance matrix. For a flat prior on $\mathcal{M}$, the nuisance parameter can be marginalized analytically, giving
\begin{equation}
\chi^2_{\rm DES,marg}=A-\frac{B^2}{C},
\end{equation}
where
\begin{align}
A &= \Delta\boldsymbol{\mu}^{T}\mathbf{C}_{\rm DES}^{-1}
\Delta\boldsymbol{\mu},\
B &= \Delta\boldsymbol{\mu}^{T}\mathbf{C}_{\rm DES}^{-1}\mathbf{1},\
C &= \mathbf{1}^{T}\mathbf{C}_{\rm DES}^{-1}\mathbf{1}.
\end{align}
Thus, the DES-SN5YR data constrain the relative distance-redshift relation while the absolute SN calibration is marginalized, avoiding the introduction of an artificial absolute calibration for $H_0$. The resulting DES contribution is then combined with the other observational datasets in the total likelihood.
For the Pantheon and Pantheon+SH0ES datasets, the absolute magnitude $M$ is explicitly included as a nuisance parameter in the fitting procedure, and its corresponding fitted value is reported in Tables. For DES-SN5YR, we do not report a separate value of $M$, since the absolute SN calibration is marginalized over in the likelihood rather than independently determined from the unanchored DES Hubble diagram.}

\textbf{Baryon Acoustic Oscillations (BAO):}  
In this analysis, we make use of a combined BAO dataset drawn from several large-scale structure surveys. These include the 6dF Galaxy Survey at an effective redshift $z_{\text{eff}}=0.106$ \cite{10.1111/j.1365-2966.2011.19250.x}, the SDSS Main Galaxy Sample at $z_{\text{eff}}=0.15$ \cite{10.1093/mnras/stv154}, the BOSS DR11 Lyman-$\alpha$ forest measurement at $z_{\text{eff}}=2.4$, the SDSS-IV eBOSS DR14 quasar observations at $z_{\text{eff}}=\{0.98,1.23,1.52,1.94\}$ \cite{eBOSS:2018yfg}, and the consensus BAO results from SDSS-III BOSS DR12 at $z_{\text{eff}}=\{0.38,0.51,0.61\}$ \cite{Alam_2017_470}. The full covariance matrices corresponding to these measurements are incorporated into the likelihood analysis to ensure a consistent statistical treatment.

The BAO observables are expressed in terms of the Hubble distance $D_H(z)$, the comoving angular diameter distance $D_M(z)$, and the volume-averaged distance $D_V(z)$, defined as
\begin{eqnarray}
D_H(z) &=& \frac{c}{H(z)},\\
D_M(z) &=& (1+z)D_A(z),\\
D_V(z) &=& \left[(1+z)^2 D_A^2(z)\frac{z}{H(z)}\right]^{1/3}, 
\end{eqnarray}

where the angular diameter distance is related to the luminosity distance through $D_A(z) = (1+z)^{-2}D_L(z)$. These quantities are combined into a set of observable vectors $F(z_i,\Theta)$, typically normalized by the sound horizon scale at the baryon drag epoch.

An important aspect of BAO analyses is the inherent degeneracy between the sound horizon $r_d$ and the Hubble constant $H_0$. This arises because BAO measurements primarily constrain the combination $c/(H_0 r_d)$, making it difficult to determine these parameters independently using BAO data alone. To break this degeneracy, one must incorporate additional cosmological information or impose external priors on $r_d$, often obtained from Cosmic Microwave Background (CMB) measurements or Big Bang Nucleosynthesis (BBN) considerations.

In this work, the comoving sound horizon at the drag epoch (\(z_d \approx 1059.94\)) is evaluated as \cite{Plank_2020}
\begin{eqnarray}
r_s(z) &=& \int_z^{\infty} \frac{c_s(\tilde{z})}{H(\tilde{z})} d\tilde{z}\\
     &=&\frac{1}{\sqrt{3}} \int_0^{1/(1+z)} \frac{da}{a^2 H(a)\sqrt{1 + \left(\frac{3\Omega_{b,0}}{4\Omega_{\gamma,0}}\right)a}}, 
\end{eqnarray}

where we adopt $\Omega_{b,0} = 0.02242$, $\Omega_{\gamma,0} = 2.469 \times 10 ^{-5}$\cite{Plank_2020}, $T_0 = 2.7255\,\text{K}$ \cite{Fixsen:2009ug}, and a fiducial value $r_{s,\text{fid}}(z_d) = 147.05\,\text{Mpc}$. The sound horizon represents the maximum comoving distance that acoustic waves can propagate through the tightly coupled photon--baryon plasma prior to recombination and consequently provides a standard ruler for BAO measurements. The BAO observations therefore constrain the expansion history through combinations of the Hubble parameter and the comoving angular-diameter distance, usually expressed in terms of $D_H(z)=c/H(z)$ and $D_M(z)$, relative to the sound horizon.

For a given cosmological model, the theoretical BAO quantities are evaluated using the corresponding expansion history $H(z)$ and compared with the observed measurements. In particular, depending on the type of BAO observable, the relevant quantities can be written as
\begin{equation}
\frac{D_H(z)}{r_s(z_d)}, \qquad
\frac{D_M(z)}{r_s(z_d)}, \qquad
\text{or}\qquad
\frac{D_V(z)}{r_s(z_d)},
\end{equation}
where
\begin{equation}
D_H(z)=\frac{c}{H(z)}, \qquad
D_M(z)=c\int_0^z\frac{dz'}{H(z')},
\end{equation}
and
\begin{equation}
D_V(z)=\left[\frac{czD_M^2(z)}{H(z)}\right]^{1/3}.
\end{equation}
Thus, the BAO data provide an important complementary constraint on the background expansion history and help break parameter degeneracies present in the supernova and cosmic-chronometer datasets.

The corresponding chi-square statistic is given by
\begin{equation}
\chi^2_{\text{BAO}}(\Theta) = \Delta F^T C^{-1}_{\text{BAO}} \Delta F,
\end{equation}
where $\Delta F = F(z_i,\Theta) - F_{\text{obs}}(z_i)$, $F(z_i,\Theta)$ denotes the theoretical BAO observable for the parameter set $\Theta$, and $F_{\text{obs}}(z_i)$ represents the corresponding observed quantity. Here, $C_{\text{BAO}}$ denotes the covariance matrix, which accounts for the correlations among the BAO measurements.

It is important to note that the sound horizon is primarily determined by the pre-recombination expansion history and therefore provides a connection between early- and late-Universe cosmology. In particular, the CMB acoustic scale constrains the combination of $r_d$ and the distance to the last-scattering surface with high precision. Consequently, an inference of a larger value of $H_0$ from late-time observations generally requires careful consideration of the assumed sound horizon. In the present analysis, we use the standard sound-horizon prescription and focus on testing modifications of the late-time expansion history within the $f(Q)$ framework. A complete assessment of the early-Universe consistency of the models would require a joint analysis with CMB likelihoods and a model-independent treatment of the $r_d$ prior, which we leave for future investigation.

The latest DESI Data Release BAO measurements provide an important extension of the earlier BAO datasets used in cosmological analyses. Unlike previous BAO compilations, DESI provides substantially improved BAO measurements over a broad redshift range, $0.1\lesssim z\lesssim4.2$, using multiple galaxy and quasar tracers. The increased redshift coverage and improved precision allow the expansion history to be tested over a wider range of cosmic epochs. In particular, the DESI measurements have revealed indications of redshift-dependent deviations from the standard $\Lambda$CDM expansion history when combined with Type Ia supernovae, making them particularly relevant for testing late-time modified-gravity scenarios. Recent studies combining DESI BAO measurements with a BBN prior have yielded an estimate of the Hubble constant around $H_0 \approx 68.51 \pm 0.58\ \mathrm{km\ s^{-1}\ Mpc^{-1}}$, independent of CMB constraints~\cite{DESI:2025zgx,qqc6-76z4}. Although this value is marginally higher than earlier Planck-based results, it remains in significant tension with local determinations from the SH0ES collaboration. This persistent discrepancy highlights the importance of BAO measurements in constraining cosmological parameters and emphasizes the need for joint analyses with complementary probes.

For each model and observational data combination, we determine the quality of the fit using the minimum chi-square statistic, $\chi^2_{\min}$, which is obtained from the maximum likelihood function $L_{\max}$ through
\begin{equation}
\chi^2_{\min} = -2\ln L_{\max}.
\end{equation}

To compare the cosmological models with the standard $\Lambda$CDM framework, we employ the Akaike Information Criterion (AIC)~\cite{Akaike_1974}, which evaluates both the goodness of fit and the number of free parameters $k$ in the model. The AIC is expressed as
\begin{equation}
\mathrm{AIC} = \chi^2_{\min} + 2k.
\end{equation}
A smaller AIC value generally indicates a more favorable model because it achieves a better balance between accurately fitting the observational data and avoiding unnecessary model complexity. Although introducing additional parameters may improve the fit, the AIC includes a penalty term to discourage over-parameterized models.

We further analyze the models using the Bayesian Information Criterion (BIC)~\cite{Schwarz_1978}, which applies a stronger penalty to models with larger parameter spaces. The BIC is defined as
\begin{equation}
\mathrm{BIC} = \chi^2_{\min} + k \ln m,
\end{equation}
where $m$ represents the total number of observational data points. Similar to the AIC, lower BIC values indicate a better model preference. However, because the penalty term depends on the sample size, the BIC suppresses overly complex models more strongly, especially for large datasets.

As a complementary Bayesian model-selection criterion, we also consider the Deviance Information Criterion (DIC) \cite{DIC_2002}, which is particularly useful for models constrained through their posterior parameter distributions. The DIC is defined as
\begin{equation}
\mathrm{DIC} = 2\overline{D(\theta)} - D(\overline{\theta}),
\end{equation}
where $D(\theta)=-2\ln L(\theta)$ denotes the deviance, $\overline{D(\theta)}$ is its posterior mean, and $D(\overline{\theta})$ is the deviance evaluated at the posterior mean of the model parameters. Unlike AIC and BIC, the DIC accounts for the effective complexity of a model through the posterior distribution rather than simply using the total number of free parameters. A lower DIC value indicates a preferred model, with differences in DIC providing a measure of the relative statistical support between competing models.

To quantify the relative performance of the considered scenarios with respect to the reference $\Lambda$CDM model, we calculate the corresponding differences in AIC, BIC, and DIC. These are defined as
\begin{equation}
\Delta \mathrm{AIC} = \mathrm{AIC}_{\rm model} - \mathrm{AIC}_{\Lambda\mathrm{CDM}}
= \Delta \chi^2_{\min} + 2\Delta k,
\end{equation}
\begin{equation}
\Delta \mathrm{BIC} = \mathrm{BIC}_{\rm model} - \mathrm{BIC}_{\Lambda\mathrm{CDM}}
= \Delta \chi^2_{\min} + \Delta k \ln m,
\end{equation}
and
\begin{equation}
\Delta \mathrm{DIC} = \mathrm{DIC}_{\rm model} - \mathrm{DIC}_{\Lambda\mathrm{CDM}}.
\end{equation}
Here, $\Delta k=k_{\rm model}-k_{\Lambda\mathrm{CDM}}$ and $\Delta\chi^2_{\min}=\chi^2_{\min,\rm model}-\chi^2_{\min,\Lambda\mathrm{CDM}}$. Lower values of $\Delta$AIC, $\Delta$BIC, and $\Delta$DIC indicate a statistical preference for the considered model relative to $\Lambda$CDM, while values close to zero indicate comparable performance. Negative values favor the considered model, whereas positive values favor the reference $\Lambda$CDM model.

The best-fit parameters for the $\Lambda$CDM model corresponding to each dataset combination are listed in Table~\ref{tableA111} of \hyperref[Appendix_I]{Appendix-I}. The statistical results for both $f(Q)$ models, including $\chi^2_{\min}$, $\Delta$AIC, $\Delta$BIC, and $\Delta$DIC, are summarized in Tables~\ref{tableA1} and~\ref{tableA11}.

\section{Analysis $f(Q)$ Models}\label{Sec:4}

The specific choice of the function \(f(Q)\) plays a central role in determining the cosmological dynamics, as it provides a purely geometric mechanism for describing the accelerated expansion of the Universe. General Relativity is recovered in the limiting case \(f(Q)=Q+2\Lambda\), while more general functional forms introduce additional contributions involving derivatives of \(f(Q)\), thereby enriching the gravitational dynamics and leading to a broader range of cosmological behaviors. Several functional forms of \(f(Q)\) have been investigated in the literature. In particular, the power-exponential model proposed in \cite{Anagnostopoulos_2021_822,SULTANA2025100422,Bohmer_2023_9,Briffa_2020_38} exhibits interesting cosmological features, including the ability to account for late-time cosmic acceleration and provide an effective description of observational data. Similarly, the power-law model studied by Paliathanasis \cite{PALIATHANASIS2025101993} leads to field equations that can be formulated as an integrable Hamiltonian system and, for specific choices of the model parameters, admits an analytical expression for the dynamical dark energy equation-of-state (EoS) parameter. In the present work, we consider two functional forms of $\Phi(Q)$ which will illustrate how different choices of $\Phi(Q)$ can accommodate the late-time cosmic expansion and potentially alleviate the $H_0$ tension, while providing viable geometric alternatives to the standard cosmological model.

\vspace{0.4cm}

\noindent

\subsection{Model I: Logarithmic \texorpdfstring{$f(Q)$}{f(Q)} Model}

In Model I, we consider a logarithmic-type functional form of $\Phi(Q)$, which is defined as
\begin{equation}
\Phi(Q)=\alpha \frac{Q^{n+1}}{Q_0^{n}}
\ln\left(\frac{Q}{Q_0}\right),
\end{equation}
where $Q=6H^2(z)$ and $Q_0=6H_0^2$. The normalization constant $\alpha$ is determined by evaluating the modified Friedmann equation at the present epoch, yielding
\begin{equation}
\alpha=\frac{\Omega_{m_0}-1}{2}.
\end{equation}

The logarithmic dependence in $\Phi(Q)$ introduces a scale-dependent correction to the standard linear form $f(Q)=Q$, with $Q_0$ serving as a reference scale associated with the present-day cosmological epoch. In particular, the correction vanishes at $Q=Q_0$, since
\begin{equation}
\Phi(Q_0)=0.
\end{equation}
The parameter $n$ controls the dependence of the modification on the nonmetricity scale and therefore determines how rapidly the deviation from the standard $f(Q)=Q$ cosmology evolves with $Q$. For convenience, we introduce the dimensionless variable
\begin{equation}
x=\frac{Q}{Q_0}.
\end{equation}
The logarithmic correction can then be written as
\begin{equation}
\Phi(Q)=\alpha Q_0 x^{n+1}\ln x.
\end{equation}
Since the full gravitational Lagrangian is given by
\begin{equation}
f(Q)=Q+\Phi(Q),
\end{equation}
the first derivative with respect to $Q$ is
\begin{equation}
f_Q
=
1+\alpha x^n
\left[
(n+1)\ln x+1
\right],
\end{equation}
and the second derivative is
\begin{equation}
f_{QQ}
=
\frac{\alpha}{Q}x^n
\left[
n(n+1)\ln x+(2n+1)
\right].
\end{equation}

These derivatives are useful for examining the theoretical viability of the model. In particular, a necessary condition for maintaining the correct sign of the gravitational kinetic term is
\begin{equation}
f_Q>0.
\end{equation}
For Model I, this condition explicitly becomes
\begin{equation}
1+\alpha
\left(\frac{Q}{Q_0}\right)^n
\left[
(n+1)\ln\left(\frac{Q}{Q_0}\right)+1
\right]>0.
\end{equation}
Therefore, the cosmological solutions should remain in the region of the parameter space where $f_Q$ is positive and does not cross zero.

We next examine the asymptotic behavior of the logarithmic correction in the high- and low-curvature regimes. Since $Q=6H^2$, the limit $Q\gg Q_0$ corresponds to the high-curvature regime, while $Q\ll Q_0$ represents the low-curvature regime. The relative contribution of the modification to the leading $Q$ term is
\begin{equation}
\frac{\Phi(Q)}{Q}
=
\alpha
\left(\frac{Q}{Q_0}\right)^n
\ln\left(\frac{Q}{Q_0}\right).
\end{equation}

For the parameter range favored by the observational analysis, Model I consistently prefers negative values of $n$. In the high-curvature limit, $Q/Q_0\rightarrow\infty$, one has
\begin{equation}
\left(\frac{Q}{Q_0}\right)^n
\ln\left(\frac{Q}{Q_0}\right)
\longrightarrow 0,
\qquad n<0.
\end{equation}
Consequently,
\begin{equation}
\boxed{
\frac{\Phi(Q)}{Q}\longrightarrow0,
\qquad Q\rightarrow\infty.
}
\end{equation}
Thus, although the logarithmic correction itself does not necessarily vanish, it becomes subdominant relative to the leading $Q$ contribution at high curvature. Furthermore,
\begin{equation}
f_Q\longrightarrow1,
\qquad
Q\rightarrow\infty,
\end{equation}
for $n<0$. Hence, Model I asymptotically approaches the standard $f(Q)\simeq Q$ behavior in the high-curvature regime. This feature is desirable because it suppresses the relative contribution of the modified-gravity correction during epochs characterized by large nonmetricity.

The low-curvature limit requires greater care. For the observationally preferred range $-1<n<0$, the correction behaves as
\begin{equation}
\Phi(Q)
\propto Q^{n+1}\ln Q.
\end{equation}
Since $n+1>0$ in this range, the correction itself satisfies
\begin{equation}
\Phi(Q)\longrightarrow0,
\qquad
Q\rightarrow0.
\end{equation}
Therefore, the modification to the background function $f(Q)$ vanishes formally in this limit. However, the behavior of its derivatives is different. In particular,
\begin{equation}
f_Q-1
\propto
Q^n\ln Q,
\end{equation}
which becomes singular as $Q\rightarrow0$ for $n<0$. Similarly, the second derivative $f_{QQ}$ may also become singular in this limit. Thus, although $\Phi(Q)$ itself tends to zero for $-1<n<0$, the logarithmic Model I does not possess a completely regular derivative structure at the strict $Q\rightarrow0$ limit. The observational constraints are obtained from finite-redshift cosmological solutions and therefore correspond to a finite range of $Q$. Consequently, the formal singular behavior at $Q\rightarrow0$ does not by itself imply a breakdown of the background solutions over the observationally relevant range. Nevertheless, it indicates that the very low-curvature regime requires a dedicated theoretical investigation before the model can be regarded as globally viable.

\begin{figure}[H]
   \includegraphics[width=8 cm]{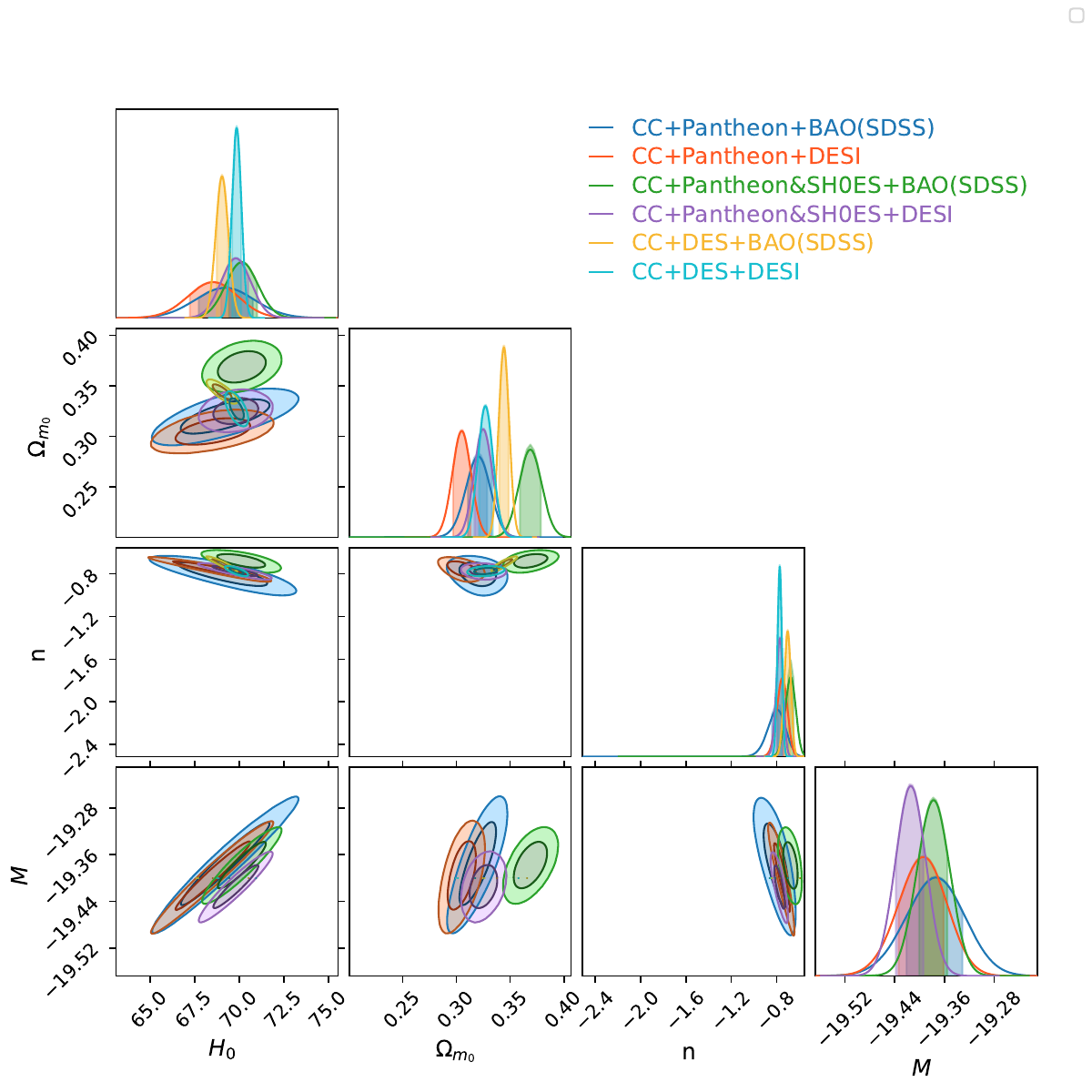}
 \caption{Contour plot for the combined dataset for the CC, Pantheon, Pantheon+SHOES, BAO(SDSS), DESI for Model-I} 
    \label{FigA1}
\end{figure}

The corresponding evolution equation for the model, where the normalized Hubble parameter is defined as $E(z)=H(z)/H_0$ takes the form
\begin{multline}
E(z)^2=\Omega_{m_0}(1+z)^3 + \alpha E(z)^{2(n+1)} \log\left(E(z)^2\right)\\
-2\alpha E(z)^{2(n+1)}\left[(n+1)\log\left(E(z)^2\right)+1\right],
\end{multline}
which is nonlinear and cannot be solved analytically, requiring numerical techniques for parameter estimation. The resulting parameter estimates are summarized in Table~\ref{tableA1}, and the corresponding confidence contours and posterior distributions are illustrated in Fig.~\ref{FigA1}. It is worth noting that in the limit $\alpha \to 0$, the model reduces to standard GR, thereby recovering the $\Lambda$CDM behavior.

\begin{figure}[H]
    \centering
    \includegraphics[width=1\linewidth]{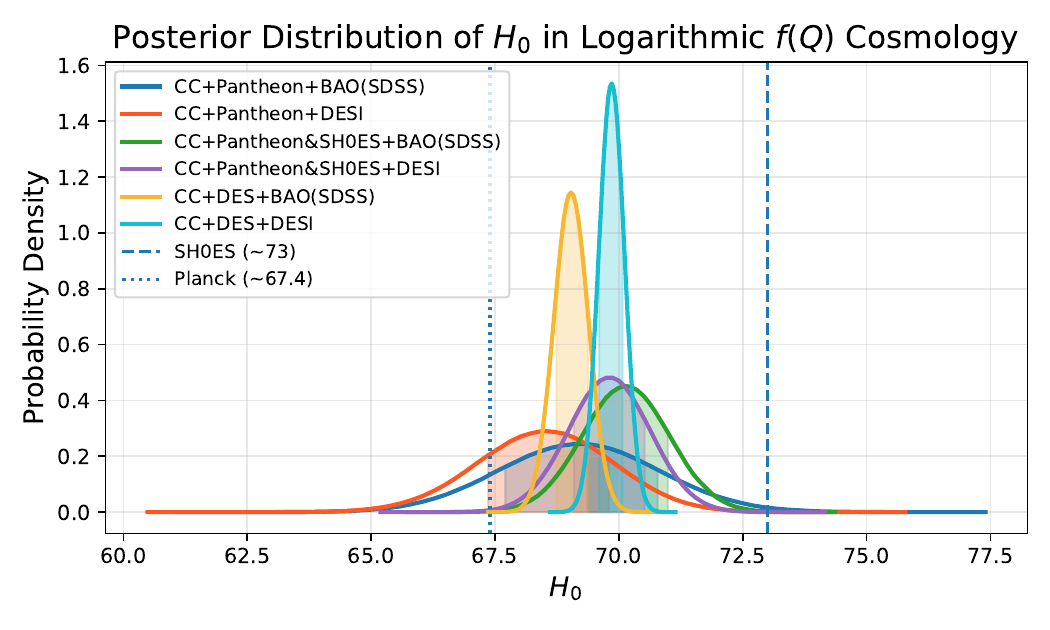}
    \caption{Posterior probability distributions of the Hubble constant $H_0$ for Model I obtained from different combinations of observational datasets.}
    \label{fig:placeholder11}
\end{figure}

For the dataset combination CC+PN+SDSS, the Hubble constant is constrained to $H_0 = 69.2^{+1.6}_{-1.7}\ \mathrm{km\ s^{-1}\ Mpc^{-1}}$, with $\Omega_{m,0} = 0.320 \pm 0.011$ and $n = -0.792^{+0.069}_{-0.079}$. Replacing SDSS with DESI (CC+PN+DESI) slightly shifts the central value to $H_0 = 68.5 \pm 1.4$ and yields tighter constraints on $\Omega_{m,0}$ and $n$, reflecting the improved precision of DESI measurements. This behavior is also evident in the posterior distributions of Fig.~\ref{fig:placeholder11}, where the DESI-based histogram appears narrower than the corresponding SDSS result, indicating reduced parameter uncertainties.

When the PN(SH0ES) dataset is included, the inferred value of the Hubble constant increases, as expected from the Cepheid calibration. Specifically, for CC+PN(SH0ES)+SDSS, we obtain $H_0 = 70.15^{+0.87}_{-0.89}$, while the CC+PN(SH0ES)+DESI combination yields $H_0 = 69.80^{+0.84}_{-0.81}$. In both cases, the uncertainties are reduced compared to the Pantheon-only results, and the parameter $n$ shifts toward less negative values, indicating a mild dependence of the model dynamics on the choice of supernova dataset. In the histogram panel, these combinations produce posterior peaks shifted toward higher values of $H_0$, bringing the results closer to the SH0ES local measurement.

A more significant tightening of constraints is observed when the DES SN5YR dataset is employed. For CC+DES+SDSS, the Hubble constant is constrained to $H_0 = 69.04 \pm 0.35$, while for CC+DES+DESI, it becomes $H_0 = 69.85 \pm 0.26$. The corresponding values of $\Omega_{m,0}$ and $n$ are also tightly constrained, with uncertainties significantly smaller than those obtained from Pantheon-based datasets. This improvement is primarily due to the larger sample size and reduced systematic uncertainties in the DES SN5YR dataset. The posterior distributions considerably sharper, demonstrating the strong constraining power of the DES SN5YR data. Across all dataset combinations, the matter density parameter $\Omega_{m,0}$ remains within the range $0.30$--$0.37$, consistent with standard cosmological estimates. The model parameter $n$ consistently takes negative values, typically around $n \sim -0.7$, indicating deviations from the $\Lambda$CDM limit while maintaining compatibility with observational data. Furthermore, the histogram distributions in Fig.~\ref{fig:placeholder11} show that the inferred values of $H_0$ generally lie between the Planck and SH0ES measurements, suggesting that the model may provide a possible route toward alleviating the $H_0$ tension.

\begin{widetext}

\begin{table}[H]
\renewcommand\arraystretch{1.5}
\centering 
{
\begin{tabular}{ c c c c c c c c c} 
\hline 
{~~~~~~~~~~\large Model I}~~~~~~~~~~~~~~ 
& ~~~~~~~~~~~~$H_0$ ~~~~~~~~~~~~~
& ~~~~~~~~~~~~$\Omega_{m,0}$~~~~~~~~~~~~ 
& ~~~~~~~~~$n$ ~~~~~~~~
& ~~~~~~~~~$M$ ~~~~~~~
& ~~~{$\chi^2_{min}$} ~~~
& ~~~{$\Delta AIC$} ~~~
& ~~~{$\Delta BIC$}~~~& ~~~{$\Delta DIC$}~~~ \\ [0.5ex] 
\hline\hline

CC+PN+SDSS 
& $69.2^{+1.6}_{-1.7}$ 
& $0.320 \pm 0.011$
& $-0.792^{+0.069}_{-0.079}$
& $-19.373^{+0.046}_{-0.048}$ 
& 1045.14 & 0.20 & 5.19 & 0.15\\
\hline
CC+PN+DESI 
& $68.5 \pm 1.4$ 
& $0.305^{+0.0086}_{-0.0088}$
& $-0.749^{+0.046}_{-0.048}$
& $-19.394^{+0.038}_{-0.040}$ 
& 1056.85 & 1.34 & 6.34 & 1.19\\
\hline

CC+PN(SH0ES)+SDSS 
& $70.15^{+0.87}_{-0.89}$ 
& $0.369^{+0.010}_{-0.011}$
& $-0.676^{+0.042}_{-0.043}$
& $-19.377^{+0.026}_{-0.027}$ 
& 1667.01 & 0.21 & 5.68 & 0.17\\
\hline
CC+PN(SH0ES)+DESI 
& $69.80^{+0.84}_{-0.81}$ 
& $0.3247^{+0.0087}_{-0.0082}$
& $-0.772^{+0.030}_{-0.031}$
& $-19.414^{+0.025}_{-0.024}$ 
& 1681.41 & 3.83 & 9.29 & 3.75\\
\hline

CC+DES+SDSS 
& $69.04 \pm 0.35$ 
& $0.3442^{+0.0047}_{-0.0049}$
& $-0.702^{+0.028}_{-0.030}$
& $\ast$ 
& 1699.50 & 1.60 & 7.14 &1.44\\
\hline
CC+DES+DESI 
& $69.85 \pm 0.26$ 
& $0.3270^{+0.0070}_{-0.0069}$
& $-0.772^{+0.019}_{-0.020}$
& $\ast$ 
& 1713.39 & 3.20 & 8.73 &3.09\\
\hline

\end{tabular}}
\caption{Model-I parameter constraints for different dataset combinations.} 

\label{tableA1}
\end{table}
\end{widetext}

\vspace{0.4cm}

\subsection{Model II: Nonlinear Saturation $f(Q)$ Model}

{In Model II, we consider a nonlinear saturation form of $\Phi(Q)$, which is designed to regulate higher-order contributions of the nonmetricity scalar and provide a controlled evolution of the cosmological dynamics across different epochs. The functional form is given by
\begin{equation}
    \Phi(Q)=\alpha Q_0
\left[
1-\left(1+\frac{Q^2}{Q_0^2}\right)^{-n}
\right],
\end{equation}
where the parameter $n$ controls the strength and rate of the nonlinear saturation, while $\alpha$ determines the overall amplitude of the modification. The normalization constant $\alpha$ is obtained by evaluating the modified Friedmann equation at the present epoch, yielding
\begin{equation}
    \alpha=
    \frac{2^n(1-\Omega_{m_0})}
    {2^n-1-2n}.
\end{equation}

The motivation for considering this functional form is to introduce a bounded nonlinear correction to the standard $f(Q)=Q$ theory.} In contrast to the logarithmic correction considered in Model I, the nonlinear structure of Model II suppresses the relative contribution of the modification at large values of the nonmetricity scalar. This provides a complementary phenomenological realization in which the late-time expansion can be modified while the correction becomes progressively less important in the high-curvature regime. Thus, Model II allows us to investigate whether a saturating modification of the nonmetricity sector can provide a viable cosmological background while remaining close to the standard gravitational behavior at sufficiently large $Q$.

For convenience, we introduce the dimensionless variable
$ x=\frac{Q}{Q_0}$.

The modification can then be written as
\begin{equation}
    \Phi(Q)=\alpha Q_0
    \left[1-(1+x^2)^{-n}\right],
\end{equation}
and the full gravitational function is
\begin{equation}
    f(Q)=Q+\Phi(Q).
\end{equation}
The first derivative of $f(Q)$ with respect to $Q$ is
\begin{equation}
    f_Q=
    1+2\alpha n x(1+x^2)^{-n-1},
\end{equation}
while the second derivative is
\begin{equation}
    f_{QQ}
    =
    \frac{2\alpha n}{Q_0}
    (1+x^2)^{-n-2}
    \left[1-(2n+1)x^2\right].
\end{equation}
These derivatives are useful for investigating the behavior of the gravitational sector and the stability of the cosmological solutions.

A necessary condition for maintaining the correct sign of the gravitational kinetic term is
\begin{equation}
    f_Q>0.
\end{equation}
For Model II, this condition takes the explicit form
\begin{equation}
    1+
    2\alpha n
    \frac{Q}{Q_0}
    \left(
    1+\frac{Q^2}{Q_0^2}
    \right)^{-n-1}
    >0.
\end{equation}
Therefore, the physically relevant cosmological solutions should remain in the region of parameter space for which $f_Q$ remains positive and does not cross zero.

The asymptotic behavior of Model II can be investigated by considering the low- and high-curvature limits. Since $Q=6H^2$, the limits $Q\ll Q_0$ and $Q\gg Q_0$ correspond respectively to the low- and high-curvature regimes. For $Q\ll Q_0$, or $x\ll1$, we can expand the nonlinear term as
\begin{equation}
    (1+x^2)^{-n}
    =
    1-nx^2
    +\frac{n(n+1)}{2}x^4
    +\mathcal{O}(x^6).
\end{equation}
Consequently,
\begin{equation}
    \Phi(Q)
    =
    \alpha Q_0
    \left[
    nx^2
    -\frac{n(n+1)}{2}x^4
    +\mathcal{O}(x^6)
    \right],
\end{equation}
or equivalently,
\begin{equation}
    \Phi(Q)
    =
    \alpha n\frac{Q^2}{Q_0}
    -
    \frac{\alpha n(n+1)}{2}
    \frac{Q^4}{Q_0^3}
    +\mathcal{O}(Q^6).
\end{equation}
Hence,
\begin{equation}
    \boxed{
    \Phi(Q)\rightarrow0,
    \qquad Q\rightarrow0,
    }
\end{equation}
and therefore
\begin{equation}
    f(Q)\rightarrow Q,
    \qquad Q\rightarrow0.
\end{equation}
Furthermore,
\begin{equation}
    f_Q
    =
    1+2\alpha n\frac{Q}{Q_0}
    +\mathcal{O}(Q^3),
\end{equation}
which gives
\begin{equation}
    \boxed{
    f_Q\rightarrow1,
    \qquad Q\rightarrow0.
    }
\end{equation}
Thus, the nonlinear saturation model has a regular low-curvature limit at the level of both $\Phi(Q)$ and $f_Q$.

The high-curvature behavior is obtained by considering $Q\gg Q_0$, or equivalently $x\gg1$. In this regime,
\begin{equation}
    (1+x^2)^{-n}
    \simeq x^{-2n}.
\end{equation}
For the observationally preferred region of Model II, $n>0$, and therefore
\begin{equation}
    x^{-2n}\rightarrow0,
    \qquad x\rightarrow\infty.
\end{equation}
Consequently,
\begin{equation}
    \Phi(Q)
    \rightarrow
    \alpha Q_0,
    \qquad Q\rightarrow\infty.
\end{equation}
Although the modification approaches a finite constant, its relative contribution to the leading nonmetricity term vanishes:
\begin{equation}
    \frac{\Phi(Q)}{Q}
    =
    \alpha\frac{Q_0}{Q}
    \left[
    1-\left(1+\frac{Q^2}{Q_0^2}\right)^{-n}
    \right]
    \longrightarrow0,
    \qquad Q\rightarrow\infty.
\end{equation}
Therefore,
\begin{equation}
    \boxed{
    \frac{\Phi(Q)}{Q}\rightarrow0,
    \qquad Q\rightarrow\infty,
    }
\end{equation}
and the leading behavior of the theory approaches
\begin{equation}
    f(Q)\simeq Q+\alpha Q_0.
\end{equation}
Since the additional term is constant while $Q$ becomes large, the relative modification to the gravitational dynamics becomes negligible. Moreover,
\begin{equation}
    f_Q
    =
    1+
    2\alpha n x^{-2n-1}
    +\mathcal{O}(x^{-2n-3}),
\end{equation}
so that
\begin{equation}
    \boxed{
    f_Q\rightarrow1,
    \qquad Q\rightarrow\infty,
    }
\end{equation}
for $n>0$. Similarly,
\begin{equation}
    f_{QQ}\rightarrow0,
    \qquad Q\rightarrow\infty.
\end{equation}
Thus, Model II possesses a regular high-curvature limit in which the modification becomes subdominant relative to the standard $Q$ contribution.

The asymptotic properties of Model II are therefore qualitatively different from those of the logarithmic Model I. In Model I, the preferred negative values of $n$ lead to a suppressed relative correction at high curvature but can result in a singular derivative structure in the formal $Q\rightarrow0$ limit. In contrast, Model II is constructed such that $\Phi(Q)$ vanishes quadratically near $Q=0$ and approaches a finite constant at large $Q$. Consequently, the nonlinear saturation form provides a more regular interpolation between the low- and high-curvature regimes.

The positivity condition $f_Q>0$ provides an important necessary viability criterion for Model II. Using the observationally constrained parameters, this condition can be evaluated along the corresponding cosmological background solutions. In particular, the observational analysis finds positive values of $n$ for all the dataset combinations considered, with the preferred value decreasing toward zero when the more precise DES and DESI data are included. This behavior indicates that the nonlinear modification becomes progressively weaker for the most tightly constrained dataset combinations.

However, the condition $f_Q>0$ is a necessary gravitational-sector viability condition and does not by itself establish complete perturbative ghost freedom. A complete stability analysis requires expanding the action to second order in scalar, vector, and tensor perturbations and examining the corresponding kinetic and gradient terms. In particular, the absence of ghost instabilities requires positive kinetic coefficients, while the absence of gradient instabilities requires non-negative squared propagation speeds. Therefore, the present analysis establishes the background asymptotic behavior and the basic condition $f_Q>0$, but does not constitute a complete proof of ghost-free perturbative stability.

The nonlinear saturation form nevertheless possesses a theoretically desirable feature in that both the $Q\rightarrow0$ and $Q\rightarrow\infty$ limits remain regular at the level of $\Phi(Q)$ and $f_Q$, with the latter approaching the standard value $f_Q=1$. A complete scalar-, vector-, and tensor-perturbation analysis, including the explicit ghost and gradient-stability conditions, is beyond the scope of the present background-level observational study and will be considered in future work.

The evolution equation for the normalized Hubble parameter $E(z)=H(z)/H_0$ takes the form
\begin{multline}
E(z)^2=\Omega_{m_0}(1+z)^3 + \alpha \\
- \alpha \left(1+E(z)^4\right)^{-n-1}
\left[E(z)^4 (4n+1)+1\right],
\end{multline}
which, similar to Model I, is nonlinear and must be solved numerically. This formulation effectively suppresses large deviations at high redshifts while allowing controlled modifications at late times.

\begin{figure}[H]
    \centering
    \includegraphics[width=8 cm]{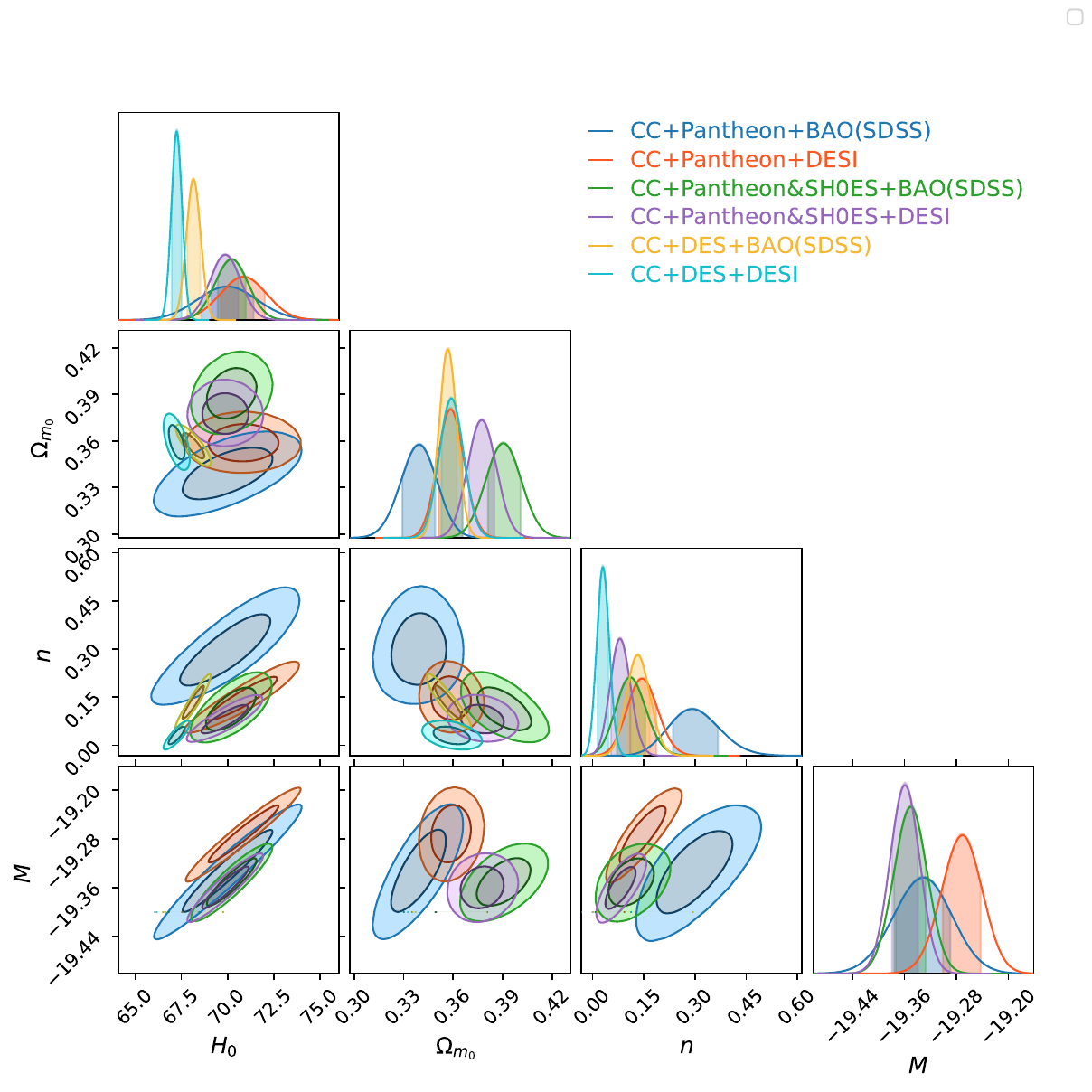}
 \caption{Contour plot for the combined dataset for the CC, Pantheon, Pantheon+SHOES, BAO(SDSS), DESI for Model-II} 
    \label{Fig11111}
\end{figure}

The constraints on the model parameters derived from various combinations of observational datasets are presented in Table~\ref{tableA11}, while the corresponding confidence contours are shown in Fig.~\ref{Fig11111}. These datasets include CC, Type Ia supernovae (Pantheon, Pantheon+SH0ES, and DES SN5YR), and BAO measurements from SDSS and DESI. The contour plots illustrate the correlations among the cosmological parameters $(H_0,\Omega_{m,0},n)$ and show that the allowed parameter space becomes progressively tighter with the inclusion of more precise datasets, particularly DESI observations.

For the CC+PN+SDSS combination, the Hubble constant is constrained to $H_0 = 70.0 \pm 1.6\ \mathrm{km\ s^{-1}\ Mpc^{-1}}$, with $\Omega_{m,0} = 0.340 \pm 0.011$ and $n = 0.292^{+0.076}_{-0.071}$. The corresponding contours in Fig.~\ref{Fig11111} show relatively broader allowed regions, especially in the $(H_0,n)$ plane, indicating weaker constraints on the nonlinear parameter $n$. When DESI data are used instead (CC+PN+DESI), the value shifts to $H_0 = 70.9^{+1.2}_{-1.3}$, accompanied by tighter constraints on $\Omega_{m,0}$ and a noticeable reduction in the parameter $n$, reflecting the impact of higher-precision BAO measurements. This improvement is clearly visible, where the confidence regions shrink considerably compared to the SDSS case.

\begin{figure}[H]
    \centering
    \includegraphics[width=1\linewidth]{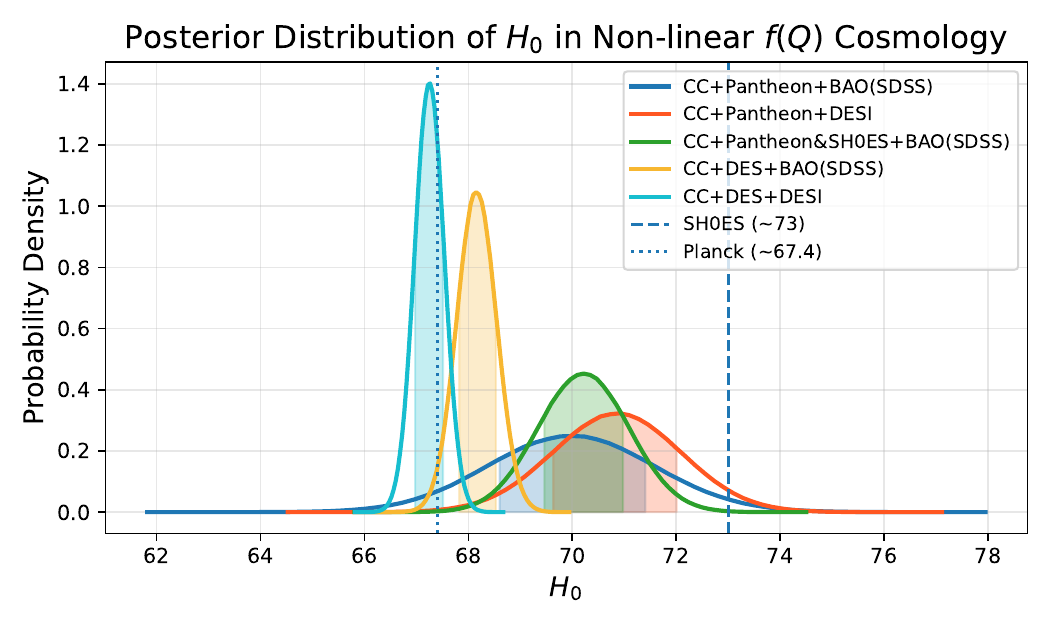}
        \caption{Posterior probability distributions of the Hubble constant $H_0$ for Model II obtained from different combinations of observational datasets.}

    \label{fig:placeholder111}
\end{figure}

Including the PN(SH0ES) dataset leads to a moderate shift in the inferred parameters. For CC+PN(SH0ES)+SDSS, we obtain $H_0 = 70.24^{+0.86}_{-0.91}$, while the CC+PN(SH0ES)+DESI combination yields $H_0 = 69.88 \pm 0.83$. In both cases, the uncertainties are reduced compared to the Pantheon-only combinations, and the parameter $n$ decreases further, suggesting weaker deviations from the standard cosmological model when local calibration data are included.

To further examine the behavior of the Hubble constant in this model, the posterior distributions of $H_0$ for all dataset combinations are presented in Fig.~\ref{fig:placeholder111}. The figure shows that the Pantheon and Pantheon+SH0ES based combinations generally favor values around $H_0 \sim 70\ \mathrm{km\ s^{-1}\ Mpc^{-1}}$, lying between the SH0ES and Planck reference measurements. This indicates that the nonlinear saturation $f(Q)$ model naturally accommodates intermediate values of the Hubble constant and partially alleviates the tension between local and early-Universe observations.

A distinct behavior is observed when the DES SN5YR dataset is considered. For CC+DES+SDSS, the Hubble constant is constrained to $H_0 = 68.15^{+0.39}_{-0.37}$, while the combination CC+DES+DESI gives $H_0 = 67.25^{+0.29}_{-0.28}$. These results exhibit significantly tighter constraints, particularly in the latter case, where the uncertainty in $H_0$ is reduced to below $0.5\%$. This trend is consistently reflected in both figures. In Fig.~\ref{Fig11111}, the confidence contours become substantially smaller and more localized, while Fig.~\ref{fig:placeholder111} shows that the DES-based datasets shift the posterior distributions toward lower values of $H_0$, especially for the CC+DES+DESI combination, whose peak lies close to the Planck measurement. Moreover, the CC+DES+DESI case produces the narrowest posterior distribution among all combinations, highlighting the strong constraining power of the combined DES and DESI observations.

The matter density parameter $\Omega_{m,0}$ remains tightly constrained across all dataset combinations, generally favoring slightly higher values compared to Model I, typically in the range $0.34$--$0.39$. The contour structure in Fig.~\ref{Fig11111} further indicates a mild correlation between $\Omega_{m,0}$ and $n$, where smaller values of $n$ are associated with slightly larger matter densities. The model parameter $n$ is consistently positive in this case, with values decreasing from $\sim 0.3$ for Pantheon-based datasets to $\sim 0.03$ for the CC+DES+DESI combination. This behavior suggests that the inclusion of increasingly precise observational data systematically drives the nonlinear saturation $f(Q)$ model closer to the $\Lambda$CDM limit, corresponding to smaller deviations from standard cosmology.

\begin{widetext}

\begin{table}[H]
\renewcommand\arraystretch{1.5}
\centering 
{
\begin{tabular}{ c c c c c c c c c} 
\hline 
{~~~~~~~~~~\large Model II}~~~~~~~~~~~~~~ 
& ~~~~~~~~~~~~$H_0$ ~~~~~~~~~~~~~
& ~~~~~~~~~~~~$\Omega_{m,0}$~~~~~~~~~~~~ 
& ~~~~~~~~~$n$ ~~~~~~~~
& ~~~~~~~~~$M$ ~~~~~~~
& ~~~{$\chi^2_{min}$} ~~~
& ~~~{$\Delta AIC$} ~~~
& ~~~{$\Delta BIC$}~~~&~~~{$\Delta DIC$}~~~ \\ [0.5ex] 
\hline\hline

CC+PN+SDSS 
& $70.0 \pm 1.6$ 
& $0.340 \pm 0.011$ 
& $0.292^{+0.076}_{-0.071}$
& $-19.330^{+0.043}_{-0.046}$ 
& 1047.23 & 2.29 & 7.28 & 2.21\\
\hline
CC+PN+DESI 
& $70.9^{+1.2}_{-1.3}$ 
& $0.3584^{+0.0081}_{-0.0078}$ 
& $0.145^{+0.045}_{-0.044}$
& $-19.271^{+0.030}_{-0.032}$ 
& 1058.54 & 3.03 & 8.03 & 2.88\\
\hline

CC+PN(SH0ES)+SDSS 
& $70.24^{+0.86}_{-0.91}$ 
& $0.390 \pm 0.011$ 
& $0.111^{+0.045}_{-0.043}$
& $-19.350^{+0.025}_{-0.026}$ 
& 1669.99 & 3.19 & 8.66 & 2.97\\
\hline
CC+PN(SH0ES)+DESI 
& $69.88 \pm 0.83$ 
& $0.3773^{+0.0088}_{-0.0085}$ 
& $0.080^{+0.030}_{-0.028}$
& $-19.359^{+0.022}_{-0.023}$ 
& 1681.92 & 4.34 & 9.80 & 4.12\\
\hline

CC+DES+SDSS 
& $68.15^{+0.39}_{-0.37}$ 
& $0.3569 \pm 0.0054$ 
& $0.133^{+0.035}_{-0.034}$
& $\ast$ 
& 1698.78 & 0.88 & 6.41 & 0.67\\
\hline
CC+DES+DESI 
& $67.25^{+0.29}_{-0.28}$ 
& $0.3591^{+0.0072}_{-0.0073}$ 
& $0.030 \pm 0.018$
& $\ast$ 
& 1712.36 & 2.17 & 7.70 & 1.98\\
\hline

\end{tabular}}
\caption{Model-II parameter constraints for different dataset combinations.} 

\label{tableA11}
\end{table}

\end{widetext}

\subsection{Comparison between Model I and Model II}
\begin{figure}[H]
    \centering
    \includegraphics[width=1\linewidth]{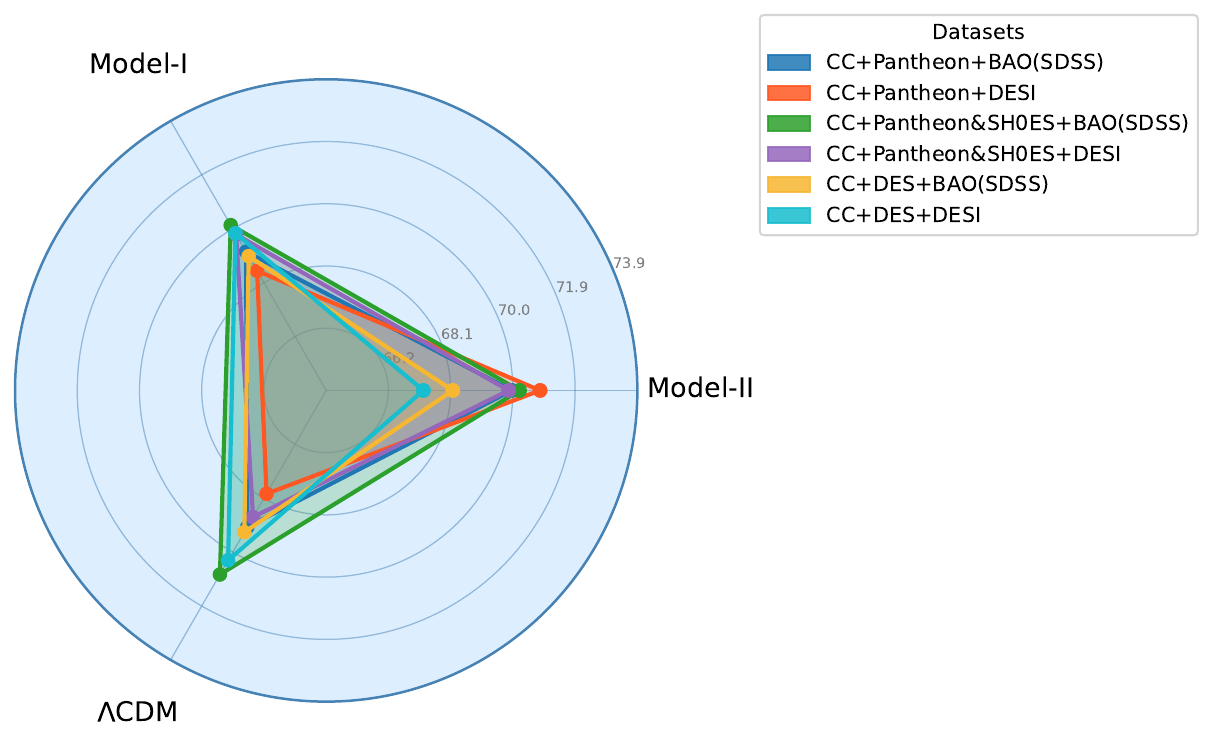}
    \caption{Plot illustrate the inferred $H_0$ values (in km s$^{-1}$ 
Mpc$^{-1}$) for Model I, Model II, and $\Lambda$CDM across six dataset 
combinations, with the radial scale ranging from $66.2$ to $73.9$ km s$^{-1}$ 
Mpc$^{-1}$.}
    \label{fig:123}
\end{figure}

A direct comparison between Model I (logarithmic) and Model II (nonlinear 
saturation) reveals both the robustness and the dataset-sensitivity of the 
inferred cosmological parameters across two distinct functional forms of $f(Q)$ 
gravity. Fig.~\ref{fig:123} displays the estimated values of $H_0$ at 
the three vertices corresponding to Model I, Model II, and $\Lambda$CDM, for 
all six dataset combinations. The overall compactness of the polygon formed by 
each dataset trace indicates that the two modified gravity models and 
$\Lambda$CDM occupy largely overlapping regions of parameter space in terms of 
$H_0$. For the CC+PN+SDSS combination, Model I gives $H_0 = 
69.2^{+1.6}_{-1.7}$ km s$^{-1}$ Mpc$^{-1}$, while Model II yields $H_0 = 
70.0 \pm 1.6$ km s$^{-1}$ Mpc$^{-1}$, a difference well within the $1\sigma$ 
uncertainties. Similarly, for CC+PN+DESI, Model I predicts $H_0 = 68.5 \pm 
1.4$ and Model II gives $H_0 = 70.9^{+1.2}_{-1.3}$. Across these baseline 
combinations, Model I consistently returns $H_0$ values that sit closer to the 
intermediate range between the Planck CMB and the SH0ES distance-ladder, whereas Model II tends to scatter slightly more, suggesting that 
the logarithmic correction provides a more stable geometric response to the 
underlying data.

When the PN(SH0ES) dataset is incorporated, both models register a 
systematic upward shift in the inferred $H_0$ regardless of the BAO counterpart 
used. For CC+PN(SH0ES)+DESI, Model I gives $H_0 = 69.80^{+0.84}_{-0.81}$ and 
Model II gives $H_0 = 69.88 \pm 0.83$, in near-perfect agreement. Likewise, 
for CC+PN(SH0ES)+SDSS the values are $H_0 \approx 70.15$ (Model I) and $H_0 
\approx 70.24$ (Model II). This behavior is clearly captured in the radar plot 
by the green trace (CC+Pantheon\&SH0ES+BAO(SDSS)), which consistently extends 
to the outermost ring near $73.9$ at the Model I vertex -- the largest $H_0$ 
excursion recorded across the entire comparison. These findings confirm that 
SH0ES calibration systematically drives $H_0$ upward independently of the 
underlying gravity model, pointing to the observational calibration as the 
dominant source of this shift.

The DES SN5YR dataset imposes the tightest constraints in both models, as 
reflected by the clustering of the yellow (CC+DES+SDSS) and cyan (CC+DES+DESI) 
traces near the center of the radar plot and by their sharply reduced 
uncertainties. For CC+DES+DESI, Model I yields $H_0 = 69.85 \pm 0.26$ and 
Model II gives $H_0 = 67.25^{+0.29}_{-0.28}$ km s$^{-1}$ Mpc$^{-1}$; the 
central values differ by approximately $2.6$ km s$^{-1}$ Mpc$^{-1}$, yet both 
carry sub-percent-level uncertainties. Here the distinction between the two 
models becomes most pronounced: Model I returns a value of $H_0 = 69.85 \pm 
0.26$ that remains well within the intermediate tension-easing range, while 
Model II yields $H_0 = 67.25^{+0.29}_{-0.28}$, which falls closer to the 
Planck estimate and therefore offers less leverage in addressing the Hubble 
tension. Similarly, for CC+DES+SDSS, Model I gives $H_0 = 69.04 \pm 0.35$ 
compared to $H_0 = 68.15^{+0.39}_{-0.37}$ from Model II, reinforcing the same 
trend. This systematic difference indicates that the logarithmic functional 
form of Model I, by virtue of its unbounded growth at large nonmetricity, is 
better equipped than the saturating correction of Model II to sustain $H_0$ at 
values capable of partially alleviating the tension with local distance-ladder 
measurements.

 In Model I, $n$ controls the power-law dependence of the logarithmic nonmetricity correction on the scale $Q/Q_0$. The preferred value $n\simeq-0.7$ implies that the relative contribution of the modification is suppressed at high curvature while allowing a scale-dependent modification at late times. In Model II, $n$ controls the rate of nonlinear saturation of the correction. For $n>0$, the modification approaches a finite value at high $Q$, while $n=0$ gives $\Phi(Q)=0$ and recovers the standard $f(Q)=Q$ limit. The preferred value decreases from approximately $n\simeq0.3$ for the CC+PN+SDSS combination toward $n\simeq0$ for the combinations involving DES and DESI, indicating a progressively weaker nonlinear modification.

The prior range of $n$ was chosen sufficiently broadly to encompass the relevant modified-gravity behavior and was fixed independently of the observational results. We adopted the flat prior and the MCMC analysis explored the complete adopted prior interval rather than being restricted to the region favored by the data. Thus, the posterior constraints on $n$ are determined by the observational likelihood rather than being imposed through the prior choice. Regions outside the adopted prior interval were not considered in the present analysis.

{    To quantify the extent to which the proposed models alleviate the Hubble tension, we calculate the tension between the inferred \(H_0\) values and the Planck and SH0ES determinations. The tension is defined as

$$
T_{A-B}=\frac{|H_0^A-H_0^B|}
{\sqrt{\sigma_A^2+\sigma_B^2}}.
$$

For the quoted Planck and SH0ES measurements the corresponding discrepancy is \(4.68\sigma\). The resulting tension values for the two \(f(Q)\) models are presented in Table~\ref{tableA1111}. Model I reduces the discrepancy with SH0ES to approximately \(2.0\)--\(3.7\sigma\) for the considered dataset combinations. Model II shows a stronger alleviation for some combinations, with the tension reaching as low as \(1.35\sigma\) for the CC+PN+DESI dataset. However, the tension remains dataset dependent, and some combinations yield a substantially larger discrepancy with SH0ES. These results indicate that the proposed \(f(Q)\) models can partially alleviate the Hubble tension.}

\begin{widetext}
    \begin{figure}[H]
    \centering
    \includegraphics[width=2.1\linewidth]{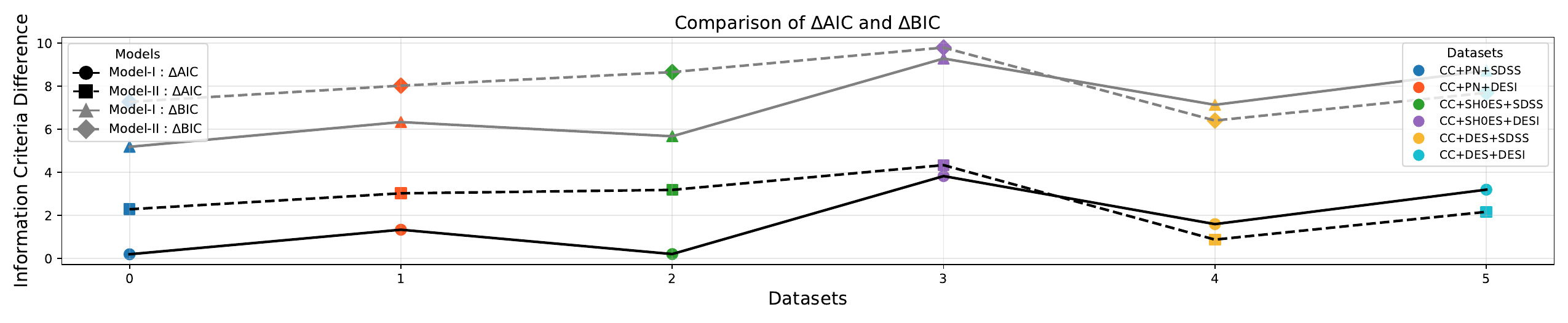}
    \caption{Variation of $\Delta$AIC and $\Delta$BIC for Model I and Model II 
relative to $\Lambda$CDM across six dataset combinations. Solid and dashed 
lines represent $\Delta$AIC and $\Delta$BIC respectively.}
    \label{fig:456}
\end{figure}
\end{widetext}

\begin{widetext}

\begin{table}[H] 
\centering 
\renewcommand{\arraystretch}{1.25} 

\label{tab} 
{\begin{tabular}{lcccc} 
\toprule 
Dataset combination~~~~~~~~~ & ~~~~~~~~~$H_0$ Model I~~~~~~~~~ & ~~~~~~~~~$T_{\rm Planck}$ / $T_{\rm SH0ES}$~~~~~~~~~ & ~~~~~~~~~$H_0$ Model II ~~~~~~~~~& ~~~~~~~~~$T_{\rm Planck}$ / $T_{\rm SH0ES}$~~~~~~~~~ \\ 
\hline
\hline
CC+PN+SDSS         & $69.2^{+1.6}_{-1.7}$       & $1.06 / 2.01$ & $70.0\pm1.6$ & $1.58 / 1.59$ \\ 
CC+PN+DESI         & $68.5\pm1.4$               & $0.79 / 2.60$ & $70.9^{+1.2}_{-1.3}$&$  2.49 / 1.35$ \\ 
CC+PN(SH0ES)+SDSS  & $70.15^{+0.87}_{-0.89}$    & $2.60 / 2.13$ & $70.24^{+0.86}_{-0.91} $&$ 2.64 / 2.07$ \\ 
CC+PN(SH0ES)+DESI  & $69.80^{+0.84}_{-0.81}$    & $2.42 / 2.42$ & $69.88\pm0.83 $&$ 2.46 / 2.37$ \\ 
CC+DES+SDSS        & $69.04\pm0.35$             & $2.37 / 3.65$ & $68.15^{+0.39}_{-0.37} $&$ 1.16 / 4.40$ \\ 
CC+DES+DESI        & $69.85\pm0.26$             & $3.64 / 2.98$ & $67.25^{+0.29}_{-0.28} $&$ 0.03 / 5.36$ \\ 
\hline
\end{tabular} }
\caption{The inferred $H_0$ values and the corresponding tension with the Planck and SH0ES measurements for Model I and Model II. The tension is expressed in units of $\sigma$.} 
\label{tableA1111}
\end{table}

\end{widetext}
The $\Delta$AIC and $\Delta$BIC values presented in Tables~\ref{tableA1} and 
\ref{tableA11}, and visualised in Fig.~\ref{fig:456}, reinforce this 
conclusion from a statistical standpoint. For Model I, $\Delta$AIC ranges from 
a minimum of $0.20$ (CC+PN+SDSS) to a maximum of $3.83$ (CC+PN(SH0ES)+DESI), 
with corresponding $\Delta$BIC values spanning $5.19$ to $9.29$. According to 
the Jeffreys scale, $\Delta$AIC $< 2$ implies substantial support relative to 
the reference model, while $2 < \Delta$AIC $< 4$ indicates a positive but 
weaker preference. Model I have substantial statistical support from the 
CC+PN+SDSS and CC+PN(SH0ES)+SDSS combinations and positive support from the 
remainder, establishing it as a statistically viable alternative to $\Lambda$CDM 
across the full range of datasets considered. For Model II, $\Delta$AIC ranges 
from $0.88$ (CC+DES+SDSS) to $4.34$ (CC+PN(SH0ES)+DESI), with $\Delta$BIC 
spanning $6.41$ to $9.80$. Although Model II
{ achieves a lower $\Delta$AIC $= 
0.88$ than Model I ($1.60$) for CC+DES+SDSS -- visible as a crossing of the 
two AIC lines at dataset index 4 in Fig.~\ref{fig:456} -- this advantage comes 
precisely in the configuration where Model II yields a lower $H_0$, offering 
less capacity to ease the Hubble tension. For CC+PN(SH0ES)+DESI (dataset index 
3), both models reach their largest information criterion differences 
($\Delta$AIC $= 3.83$ and $4.34$; $\Delta$BIC $= 9.29$ and $9.80$ for Models 
I and II respectively), with Model I retaining the smaller penalty throughout. 
The BIC curves lie systematically above the AIC curves for both models across 
all dataset combinations, reflecting the expected large-$n$ penalty, while the 
two models track each other closely in their information criterion profiles. 
We also consider the Deviance Information Criterion (DIC), which accounts for both the goodness of fit and the effective complexity of the model. The $\Delta$DIC values for Model I range from $0.15$ to $3.75$, whereas those for Model II range from $0.67$ to $4.12$. Model I gives the smallest $\Delta$DIC for CC+PN+SDSS ($\Delta \mathrm{DIC}=0.15$) and remains below $2$ for the CC+PN+SDSS, CC+PN+DESI, CC+PN(SH0ES)+SDSS, CC+DES+SDSS, and CC+DES+DESI combinations, indicating a comparatively small DIC difference from $\Lambda$CDM. Model II also shows competitive DIC values for CC+PN+SDSS and CC+DES+SDSS, with $\Delta \mathrm{DIC}=2.21$ and $0.67$, respectively, but reaches its largest difference of $4.12$ for CC+PN(SH0ES)+DESI. Taken together, the AIC, BIC, and DIC results show that Model I is generally more competitive with $\Lambda$CDM than Model II across the considered dataset combinations. In particular, Model I simultaneously provides relatively small information-criterion differences and higher $H_0$ values for most datasets, making the logarithmic correction a more consistent candidate for partially alleviating the Hubble tension. Nevertheless, the statistical preference remains dataset dependent, and neither model is strongly preferred over $\Lambda$CDM for all observational combinations.}

\section{Results and Discussions}\label{Sec:5} 

In this work, we investigated the cosmological implications of two representative $f(Q)$ gravity models within the framework of symmetric teleparallel gravity using a comprehensive set of late-time observational datasets, including CC, Type Ia supernova compilations (Pantheon, Pantheon+SH0ES, and DES SN5YR), and baryon acoustic oscillation (BAO) measurements from SDSS and DESI.

For the logarithmic Model-I, the inferred values of the Hubble constant generally lie in the range $H_0 \sim 68.5$--$70.2~\mathrm{km\,s^{-1}\,Mpc^{-1}}$, depending on the adopted dataset combination. In particular, the inclusion of the Pantheon+SH0ES compilation systematically shifts the preferred value of $H_0$ toward higher values, bringing the results closer to the local SH0ES determination. The DESI datasets significantly improve the precision of the constraints compared to SDSS, producing narrower confidence contours and posterior distributions. Moreover, the DES SN5YR dataset yields the tightest parameter constraints, with uncertainties reduced to sub-percent levels in some combinations. Across all datasets, the model parameter $n$ remains negative, indicating deviations from the $\Lambda$CDM limit while preserving consistency with cosmological observations. Importantly, the posterior distributions show that the logarithmic correction naturally favors intermediate values of $H_0$ between the Planck and SH0ES measurements, suggesting that Model-I can partially alleviate the Hubble tension.

For the nonlinear saturation Model-II, the cosmological constraints exhibit a somewhat different behavior. The Pantheon and Pantheon+SH0ES combinations again favor values around $H_0 \sim 70~\mathrm{km\,s^{-1}\,Mpc^{-1}}$, but the inclusion of DES and DESI data systematically drives the inferred Hubble constant toward lower values closer to the Planck estimate. In particular, the CC+DES+DESI combination yields $H_0 = 67.25^{+0.29}_{-0.28}~\mathrm{km\,s^{-1}\,Mpc^{-1}}$, which is highly consistent with early-Universe measurements but less effective in addressing the local Hubble tension. The nonlinear parameter $n$ remains positive and gradually approaches zero with increasingly precise datasets, indicating that the model tends toward the $\Lambda$CDM limit when confronted with stronger observational constraints. This behavior suggests that the saturation mechanism suppresses large deviations from standard cosmology at late times.

A direct comparison between the two models reveals that Model-I generally provides a more favorable framework for easing the Hubble tension. The logarithmic correction consistently predicts intermediate values of $H_0$ that remain statistically compatible with both early- and late-Universe measurements, whereas Model-II tends to converge toward lower values favored by Planck. This distinction becomes especially pronounced for the DES+DESI combinations, where Model-I retains $H_0$ values near $70~\mathrm{km\,s^{-1}\,Mpc^{-1}}$ while Model-II approaches the standard $\Lambda$CDM prediction. These results indicate that the logarithmic form of the nonmetricity correction introduces a more flexible late-time expansion history capable of sustaining larger Hubble values without significantly degrading the fit to observational data.

The statistical comparison based on the AIC, BIC and DIC further supports these conclusions. Both models remain statistically competitive with the $\Lambda$CDM cosmology, with relatively small values of $\Delta \mathrm{AIC}$, $\Delta \mathrm{BIC}$ and $\Delta \mathrm{DIC}$ across all dataset combinations. However, Model-I generally exhibits lower information criterion differences than Model-II, particularly for the Pantheon-based datasets, indicating a better balance between model complexity and goodness of fit. Although Model-II performs slightly better for the CC+DES+SDSS combination, this occurs precisely in the regime where the inferred value of $H_0$ shifts closer to the Planck estimate, reducing its effectiveness in addressing the Hubble tension.

For completeness, we also analyzed the standard $\Lambda$CDM model using the same observational datasets. The results show that the inferred value of $H_0$ remains highly sensitive to the adopted late-time observations. Pantheon-only combinations yield values close to the Planck estimate, while Pantheon+SH0ES combinations shift the preferred $H_0$ upward toward the local SH0ES measurement. DESI data consistently provide tighter constraints than SDSS, highlighting the increasing precision of recent BAO observations. Overall, the $\Lambda$CDM framework continues to provide an excellent fit to the available data, although the dependence of $H_0$ on dataset selection demonstrates that the Hubble tension remains unresolved within the standard cosmological scenario.

{In summary, the two \(f(Q)\) models provide viable descriptions of the observed late-time expansion history, while exhibiting distinct effects on the inferred value of \(H_0\). The logarithmic model shows a comparatively stronger tendency to shift the inferred \(H_0\) toward the local SH0ES measurement, whereas the nonlinear model provides a consistent fit with a weaker departure from the standard \(f(Q)=Q\) behavior for some dataset combinations. These results indicate that modifications of the nonmetricity sector can influence the late-time expansion rate and consequently affect the value of the Hubble constant inferred from cosmological observations. The statistical analysis further shows that the relative performance of the two models depends on the combination of datasets, reflecting the complementary constraining power of the different observational probes. In particular, the inclusion of recent BAO measurements provides additional constraints on the expansion history and helps distinguish between the different functional forms of \(f(Q)\). The combined observational results therefore support the viability of these models as phenomenological alternatives to \(\Lambda\)CDM, while also indicating that their ability to alleviate the \(H_0\) tension is not sufficient to claim a complete resolution. The logarithmic model appears particularly promising in this respect, as it provides a closer agreement with the local determination of \(H_0\) while maintaining compatibility with the considered cosmological datasets. Further improvements in observational precision will be important to determine whether this preference persists and whether modified late-time dynamics are genuinely required.}

\section*{Acknowledgements} BM acknowledges IUCAA, Pune (India) for providing support in the form of an academic visit during which part of this work is accomplished. The authors are thankful to the honorable referees for their constructive comments and suggestions for the improvement of the paper.

\section*{Appendix-I}\label{Appendix_I}

The parameter constraints for the $\Lambda$CDM model obtained from different combinations of CC, Pantheon/Pantheon+SH0ES, DES, SDSS, and DESI datasets are summarized in Table~\ref{tableA111}. The analysis shows that the estimated value of the Hubble constant $H_0$ depends sensitively on the choice of observational datasets. For the combinations involving Pantheon and BAO data, the obtained values of $H_0$ lie in the range $67.9$--$69.1~\mathrm{km,s^{-1},Mpc^{-1}}$, remaining broadly consistent with the $\textit{Planck}$ CMB estimate. In contrast, the inclusion of the Pantheon+SH0ES compilation shifts the preferred value of $H_0$ toward larger values, reaching $70.81 \pm 0.72~\mathrm{km,s^{-1},Mpc^{-1}}$ for the CC+PN(SH0ES)+SDSS dataset combination, thereby reducing the tension with the local SH0ES measurement.

\begin{figure}[H]
    \centering
    \includegraphics[width=8 cm]{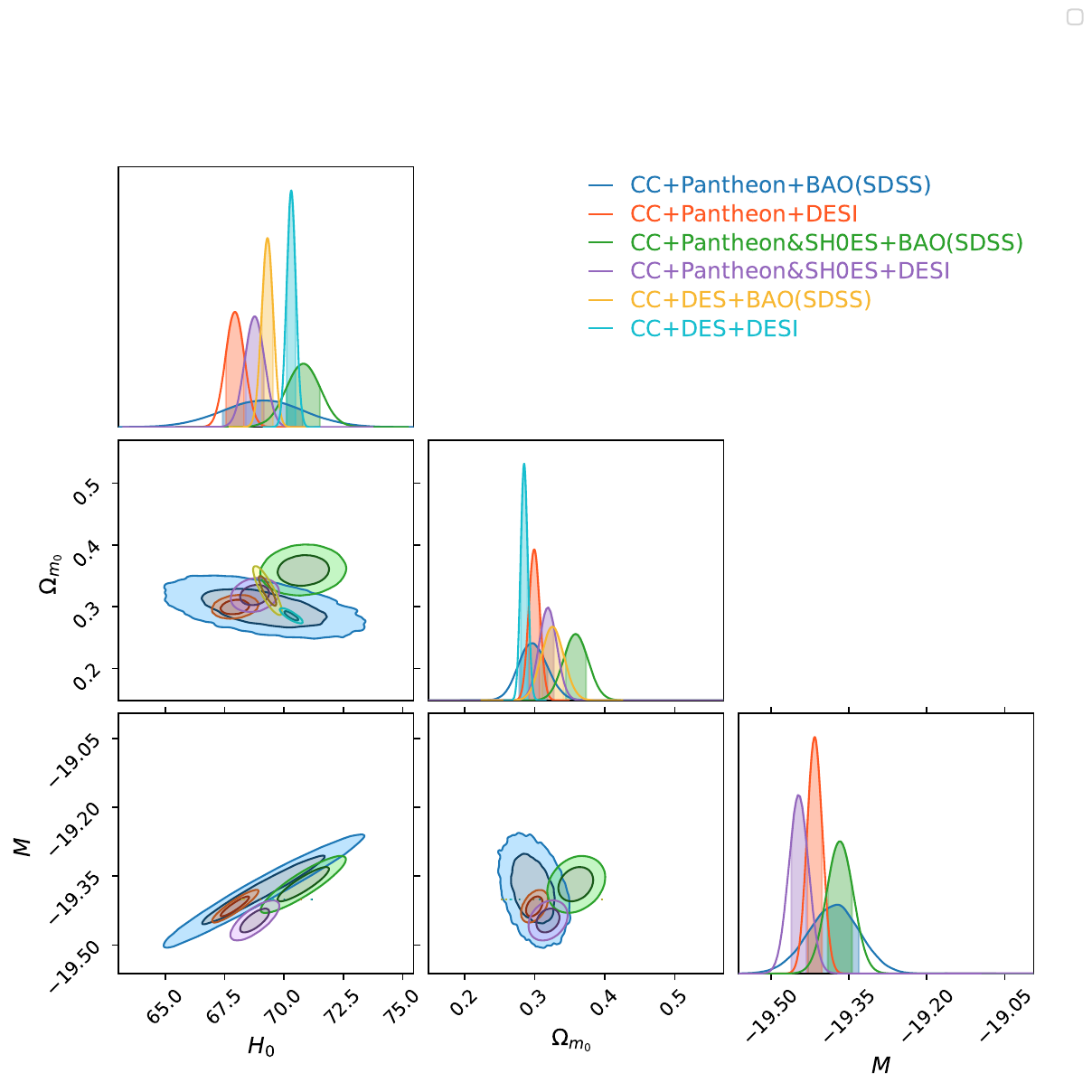}\\
 \caption{Contour plot for the combined dataset for the CC, Pantheon, Pantheon+SH0ES, BAO(SDSS), DESI for $\Lambda$CDM} 
    \label{Fig101}
\end{figure}

The matter density parameter $\Omega_{m,0}$ is tightly constrained for all dataset combinations, with values lying around $\Omega_{m,0}\sim0.28$--$0.36$. In particular, the DESI combinations provide significantly tighter bounds compared to the SDSS datasets, indicating the improved constraining power of the recent DESI BAO measurements. The absolute magnitude parameter $M$ is also well constrained in the Pantheon-based analyses, with best-fit values around $M\approx -19.4$.

The minimum chi-square values and the corresponding information criteria (AIC, BIC and DIC) indicate that all dataset combinations provide statistically acceptable fits within the $\Lambda$CDM framework. However, the DESI datasets generally yield slightly lower relative information criteria compared to SDSS, suggesting a marginally better consistency with the $\Lambda$CDM cosmology. Overall, the results demonstrate that the standard $\Lambda$CDM model continues to provide an excellent description of the current cosmological observations, while the inferred value of $H_0$ still exhibits a noticeable dependence on the adopted late-time datasets (Fig.\ref{Fig101}).

\begin{figure}[H]
    \centering
    \includegraphics[width=1\linewidth]{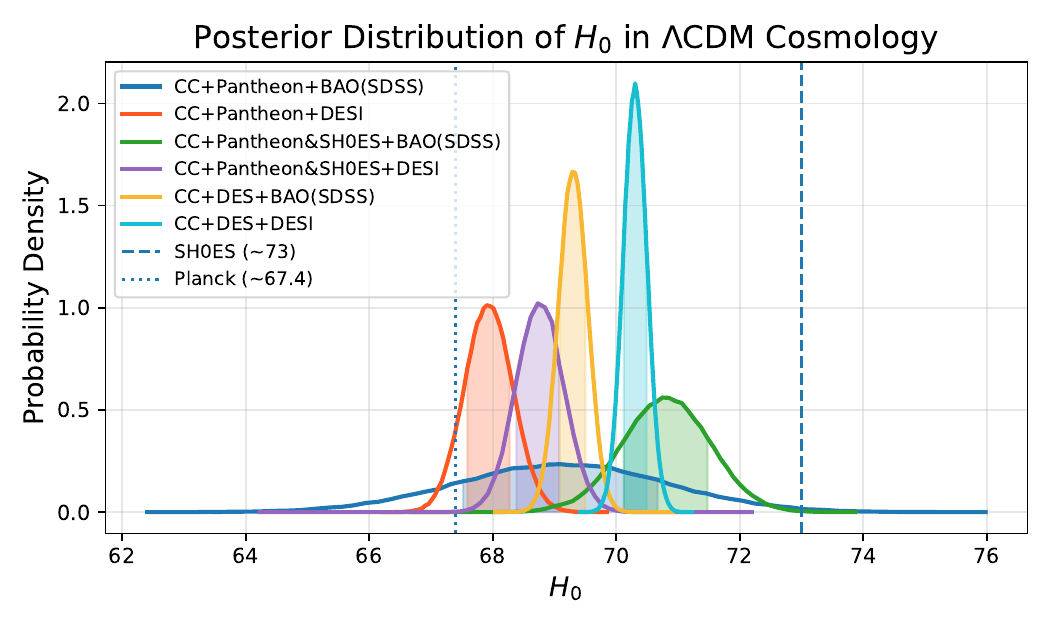}
    \caption{Posterior probability distributions of the Hubble constant $H_0$ for $\Lambda$CDM obtained from different combinations of observational datasets.}
    \label{fig:placeholderM}
\end{figure}

The posterior distributions shown in  further illustrate the dataset dependence of $H_0$ (Fig.\ref{fig:placeholderM}). The Pantheon+SH0ES combinations favor higher values of the Hubble constant closer to the SH0ES estimate ($\sim73~\mathrm{km,s^{-1},Mpc^{-1}}$), whereas the Pantheon and DES-based combinations remain closer to the $\textit{Planck}$ prediction ($\sim67.4~\mathrm{km,s^{-1},Mpc^{-1}}$). Similarly, the combined contour plots in  show the correlations among $H_0$, $\Omega_{m,0}$, and $M$, demonstrating that DESI data lead to narrower confidence contours and therefore tighter cosmological parameter constraints.

\begin{widetext}

\begin{table}[H]
\renewcommand\arraystretch{1.5}
\centering 
{
\begin{tabular}{ c c c c c c c c} 
\hline 
{~~~~~~~~~~\large $\Lambda$CDM}~~~~~~~~~~~~~~ 
& ~~~~~~~~~~~~$H_0$ ~~~~~~~~~~~~~
& ~~~~~~~~~~~~$\Omega_{m,0}$~~~~~~~~~~~~ 
& ~~~~~~~~~~~~$M$ ~~~~~~~~~~~~
& ~~~~~~~{$\chi^2_{min}$} ~~~~~~~
& ~~~~~~~{AIC} ~~~~~~~
& ~~~~~~~{BIC}~~~~~~~ & ~~~~~~~{DIC}~~~~~~~ \\ [0.5ex] 
\hline\hline

CC+PN+SDSS 
& $69.1 \pm 1.7$ 
& $0.297^{+0.020}_{-0.021}$ 
& $-19.370^{+0.042}_{-0.057}$ 
& 1046.94 & 1052.94 & 1067.92 &1050.94\\

\hline

CC+PN+DESI 
& $67.92^{+0.40}_{-0.39}$ 
& $0.2994^{+0.0076}_{-0.0079}$ 
& $-19.415^{+0.014}_{-0.015}$ 
& 1057.51 &  1063.51& 1078.50 &1061.51 \\

\hline

CC+PN(SH0ES)+SDSS 
& $70.81 \pm 0.72$ 
& $0.359^{+0.016}_{-0.017}$ 
& $-19.368^{+0.026}_{-0.024}$ 
& 1668.80 & 1674.80 & 1691.19 &1672.80\\

\hline

CC+PN(SH0ES)+DESI 
& $68.77^{+0.39}_{-0.41}$ 
& $0.3176^{+0.0120}_{-0.0098}$ 
& $-19.447^{+0.018}_{-0.017}$ 
& 1679.58 & 1685.58 & 1701.97 &1685.58\\

\hline

CC+DES+SDSS 
& $69.31^{+0.23}_{-0.25}$ 
& $0.325 \pm 0.016$ 
& $\ast$ 
& 1699.90 & 1703.90 & 1714.97 &1703.90\\

\hline

CC+DES+DESI 
& $70.30^{+0.20}_{-0.19}$ 
& $0.2848^{+0.0051}_{-0.0049}$ 
& $\ast$ 
& 1712.19 & 1716.19 & 1727.26 &1716.19\\
\hline

\end{tabular}}
\caption{$\Lambda$CDM parameter constraints for different dataset combinations.} 

\label{tableA111}
\end{table}

\end{widetext}

\section*{References}
 \bibliographystyle{utphys}
\bibliography{references}


\end{document}